\documentclass{article}

\usepackage{PRIMEarxiv}

\usepackage{amsfonts}       % blackboard math symbols
\usepackage{amsmath}        % math formulas
\usepackage{booktabs}       % professional-quality tables
\usepackage{fancyhdr}       % header
\usepackage[T1]{fontenc}    % use 8-bit T1 fonts
\usepackage{graphicx}       % graphics
\usepackage{hyperref}       % hyperlinks
\usepackage[utf8]{inputenc} % allow utf-8 input
\usepackage{lipsum}
\usepackage{makecell}
\usepackage{microtype}      % microtypography
\usepackage{multirow}
\usepackage[super,sort&compress,comma]{natbib}
\usepackage{nicefrac}       % compact symbols for 1/2, etc.
\usepackage[abs]{overpic}
\usepackage{subcaption}
\usepackage{textgreek}
\usepackage{url}            % simple URL typesetting

\graphicspath{{media/}}     % organize your images and other figures under media/ folder

\title{Insights into \NbC{2} and \NbCO{2} as high-performance anodes for sodium- and lithium-ion batteries: An \textit{ab initio} investigation
}

\author{
  Nishat Sultana \\
  Department of Mechanical and Materials Engineering \\
  University of Cincinnati \\
  2901 Woodside Drive, Cincinnati, OH, 45219\\
  \texttt{sultannt@mail.uc.edu} \\
  \And
  Abdullah A. Amin \\
  Department of Mechanical and Aerospace Engineering \\
  University of Dayton \\
  300 College Park Drive, Dayton, OH 45469\\
  \texttt{aamin1@udayton.edu} \\
  \And
  Eric J. Payton, Woo Kyun Kim \\
  Department of Mechanical and Materials Engineering \\
  University of Cincinnati \\
  2901 Woodside Drive, Cincinnati, OH, 45219\\
  \texttt{\{paytonej, kimwu\}@ucmail.uc.edu} \\
}

\newcommand{\NbC}[1]{Nb\textsubscript{#1}C}
\newcommand{\NbCO}[1]{Nb\textsubscript{#1}CO\textsubscript{#1}}
\newcommand{\NbCLi}[1]{Nb\textsubscript{#1}CLi}
\newcommand{\NbCNa}[1]{Nb\textsubscript{#1}CNa}

\begin{document}
\maketitle

\begin{abstract}
  In this study, we employ first-principles density functional theory (DFT) calculations to investigate the electrochemical properties of \NbC{2} and \NbCO{2} MXenes as potential anode materials for sodium-ion (SIBs) and lithium-ion batteries (LIBs). Our findings reveal that Li and Na intercalation primarily modifies the electronic properties of \NbC{2} without inducing significant structural distortions, as indicated by Raman intensity variations. Adsorption energy calculations show that the T4 and H3 sites are the most favorable for metal intercalation, with \NbCO{2} exhibiting stronger adsorption due to oxygen functionalization.
  We find that \NbC{2} offers lower diffusion barriers, especially for Na ions, making it a promising candidate for fast-charging SIBs. In contrast, \NbCO{2} enhances charge retention through stronger electrostatic interactions but introduces higher migration resistance. Electronic structure analysis confirms the metallic nature of both MXenes, ensuring efficient electron transport. Open-circuit voltage (OCV) calculations indicate that \NbCO{2} exhibits higher OCV values than \NbC{2}, highlighting the role of surface functionalization in tuning electrochemical performance.
  Our study suggests that, while Li-based systems achieve slightly higher theoretical capacities, Na-based systems exhibit comparable performance, reinforcing the viability of sodium-ion batteries as a cost-effective alternative. Overall, our results demonstrate that \NbC{2} is better suited for rapid ion transport, whereas \NbCO{2} offers enhanced charge retention. These insights provide a foundation for the optimization of MXene-based electrodes for next-generation high performance energy storage applications.
\end{abstract}

% keywords can be removed
\keywords{Sodium-Ion batteries \and Lithium-Ion batteries \and Density Functional Theory \and MXenes \and Anodes \and First-principles calculations}

\begin{figure*}
  \centering
  \begin{subfigure}{0.36\textwidth}
    \centering
    \caption*{\raisebox{1.5\height}{\hspace{25mm}\textbf{(a)}}} % Adjust position
    \vspace{2mm}
    \includegraphics[width=\textwidth]{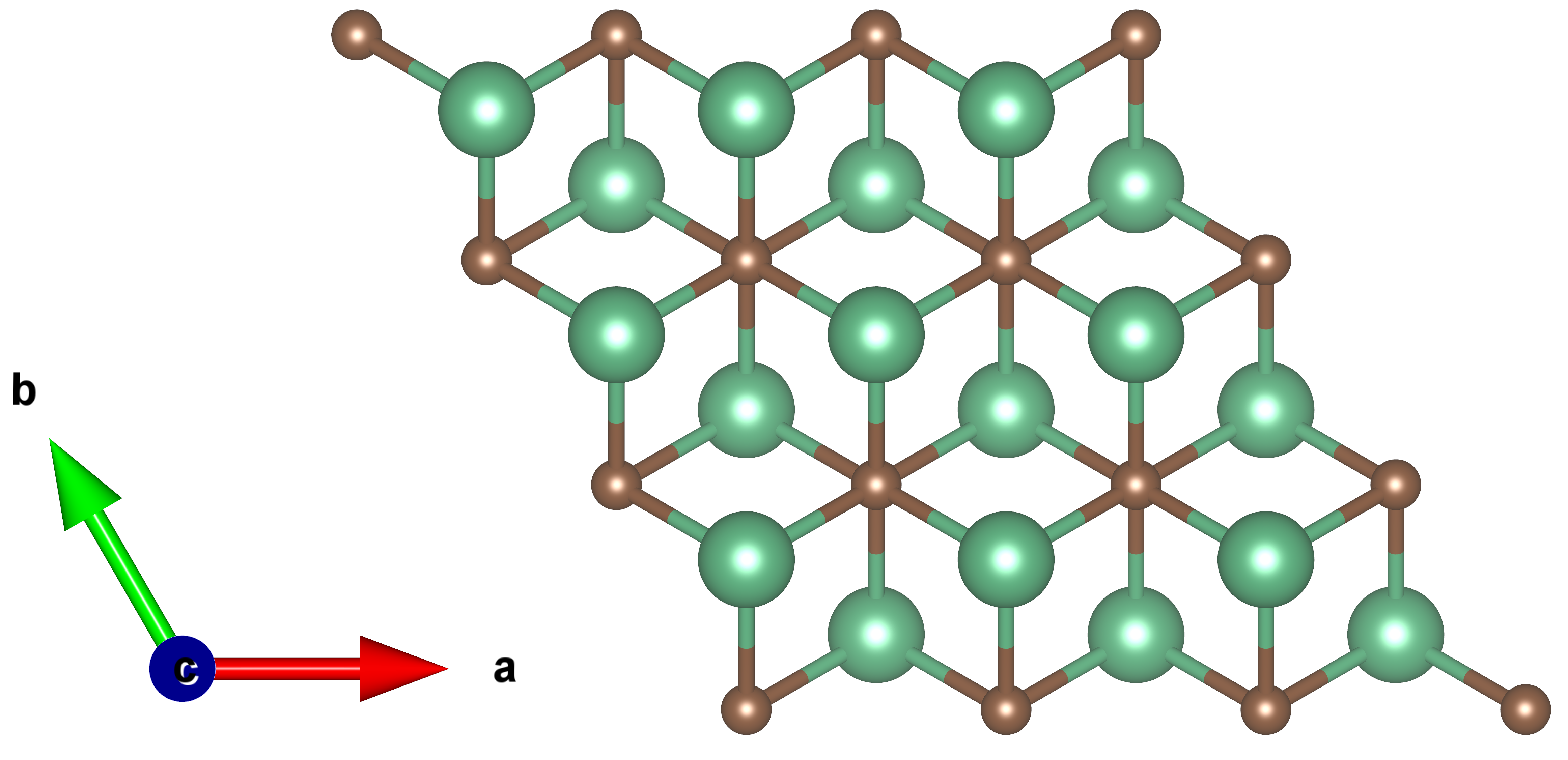}
    % \caption*{(a)}
  \end{subfigure}
  \begin{subfigure}{0.27\textwidth}
    \centering
    \caption*{\raisebox{1.5\height}{\hspace{25mm}\textbf{(b)}}} % Adjust position
    \vspace{-2mm}
    \includegraphics[width=\textwidth]{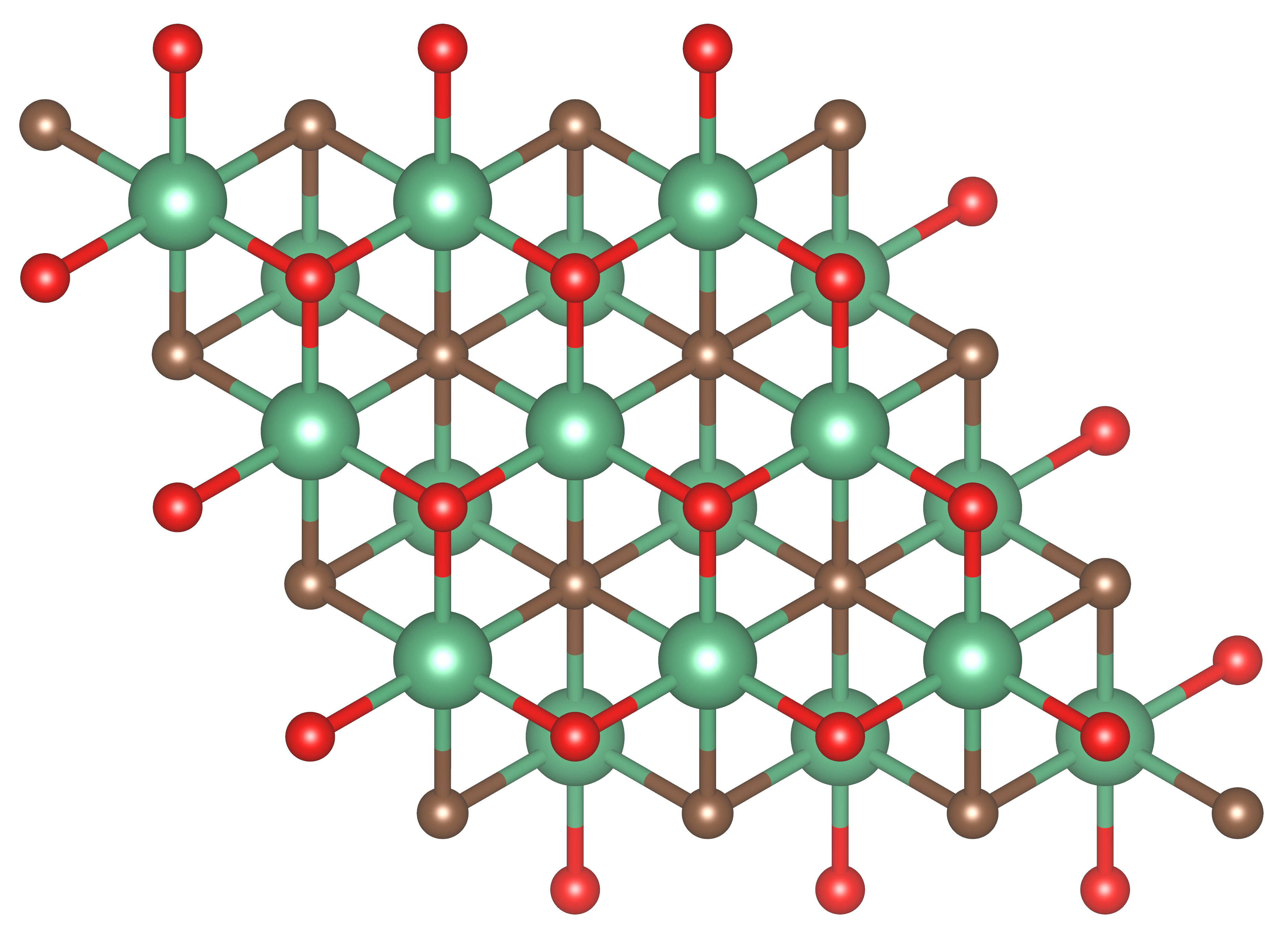}
  \end{subfigure}
  \hspace{6mm}
  \begin{subfigure}{0.27\textwidth}
    \centering
    \caption*{\raisebox{1.5\height}{\hspace{25mm}\textbf{(c)}}} % Adjust position
    \vspace{2mm}
    \includegraphics[width=\textwidth]{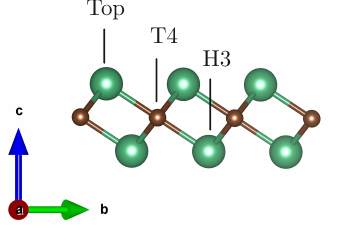}
  \end{subfigure}
  \caption{Atomistic models for the (a) \NbC{2} and (b) \NbCO{2} MXenes. (c) represent the possible adsorption sites for metal atoms.}
  \label{fgr:atom_model_Nb2C_Nb2CO2}
\end{figure*}

\section{Introduction}
The escalating global energy demand---from electric vehicles (EVs) and grid-scale battery systems to portable electronics devices---necessitates the advancement of efficient and cost-effective energy storage solutions\cite{waseemEnergyStorageTechnology2025}. As renewable energy adoption grows, technologies like solar and wind power further emphasize the need for reliable batteries capable of storing and delivering energy on demand\cite{sayedRenewableEnergyEnergy2023}. In this context, lithium-ion batteries (LIBs) have become the dominant energy storage technology due to their high energy density, long cycle life, and relatively fast charge-discharge rates\cite{liuCurrentFutureLithiumion2021}. Their widespread use extends across consumer electronics, electric mobility, and renewable energy storage, enabling advancements in everything from smartphones and laptops to EVs and grid stabilization systems. However, despite their advantages, the increasing demand for LIBs has raised concerns regarding lithium's limited global supply, uneven distribution, and rising extraction costs\cite{vinayakCircularEconomiesLithiumion2024}. These challenges have driven the search for alternative battery technologies that can offer comparable performance at a lower cost. 

Sodium-ion batteries (SIBs) have emerged as a promising alternative due to their natural abundance and relative lower cost than lithium\cite{wanisonEngineeringAspectsSodiumion2024, abrahamHowComparableAre2020}. While in general, SIBs can be 10–20\% more cost-effective than LIBs\cite{abrahamHowComparableAre2020}, their cost advantage can reach up to 25\% on some occasions\cite{rudolaOpportunitiesModeraterangeElectric2023}. Although Na has a larger ionic radius, leading to differences in electrochemical behavior, recent advancements in electrode materials have demonstrated that SIBs can achieve competitive and sometimes superior performance in terms of energy density\cite{kaliyappanConstructingSafeDurable2020}, cycle life\cite{fanResearchProgressHard2024,dingAdvancesMnBasedElectrode2023}, low temperature operation\cite{mengStrategiesImprovingElectrochemical2022}, thermal stability\cite{sunLonglifeNarichNickel2021}, and safety\cite{gaoEnormousPotentialSodium2024}. Central to these improvements is the development of efficient anode materials that offer superior reaction kinetics\cite{tianHeterogeneousStructuredSnO22023}, high conductivity\cite{liMXenebasedAnodeMaterials2024}, structural stability\cite{zhangResearchProgressPreparation2025}, and low ion diffusion barriers\cite{dandiguntaAGraphyneUltralowDiffusion2024}, which are key to unlocking the full potential of SIBs as a cost-effective and sustainable energy storage solution\cite{wanFastChargingAnodeMaterials2024}. Previous studies have shown that SIBs require greater structural and lattice stability in electrode materials, making many materials that excel in lithium storage unsuitable for sodium storage applications\cite{kumarHighPerformanceSodiumIonBatteries2025}. For example, in commercial lithium-ion batteries (LIBs), graphite is widely used as an anode material, but it fails to accommodate sodium ions effectively due to the insufficient interlayer spacing of graphite ($\sim$0.334 nm) to accommodate the larger $Na^{+}$ ions, making it unsuitable for sodium-ion batteries (SIBs)\cite{wenExpandedGraphiteSuperior2014}. In addition, limited active sites of hard carbon for $Na^{+}$ ions contributes to reduced capacity and rate performance\cite{zhangDefectEngineeringElectrode2020}. Moreover, many anode materials suffer from poor cycling stability\cite{liuHighPerformanceAnodes2023}, low specific capacity\cite{rizaReviewAnodeMaterials2024}, and low initial Coulombic efficiency (ICE)\cite{fangTransitionMetalOxide2020}, which hinder the practical implementation of SIBs\cite{kumarHighPerformanceSodiumIonBatteries2025}. As a result, significant efforts and attention have been directed toward identifying stable anode materials with desirable electrochemical properties, as they play a crucial role in advancing the next generation of renewable energy technologies. 

In this aspect, MXenes (M=transition metal, X=Carbide/nitride layers), a novel category of two-dimensional transition metal compounds, typically derived from MAX (M=transition metal, A=A-group element, X= carbon/nitrogen) phases through selective chemical etching of the A-layer, have emerged as promising anode materials for energy storage devices \cite{zhuRecentAdvancesMXenebased2022}. Their structural design, excellent electrical conductivity, expansive surface area, hydrophilicity\cite{anasori2DMetalCarbides2017}, and adaptable surface chemistry make them well-suited for energy storage applications\cite{lokhandeProspectsMXenesEnergy2022}. MXenes are typically represented as $M_{n+1}X_{n}T_{x}$, where $T_{x}$ represents surface functional groups (-OH, -O, -F) introduced during the etching process\cite{pogorielovMXenesNewClass2021}. Since, the discovery of this outstanding material, significant studies have been conducted on different Mxenes such as $Ti_{2}C$, $Ti_{3}C_{2}$, $Mo_{2}C$, $Nb_{2}C$, and $V_{2}C$   to explore their  efficacy as potential anode materials for energy storage\cite{EngineeringTi3C2MXeneSurface, ElucidatingChargeStorage,ponnalagarRecentAdvancesFuture2024, ponnalagarRecentProgressTwodimensional2023,behlRecentDevelopmentsV2C2024}. Among various MXenes, niobium-based \NbC{2} and its oxygen-functionalized derivative, \NbCO{2}\cite{naguibNewTwoDimensionalNiobium2013}, have garnered significant attention as promising electrode material in energy storage devices  due to their metallic conductivity, excellent mechanical stability, and high theoretical capacities\cite{ponnalagarRecentProgressTwodimensional2023}. The presence of surface terminations, such as oxygen (-O), plays a critical role in modulating the adsorption, diffusion, and electrochemical performance of intercalating ions, influencing the overall battery performance. Despite these advantages, a comprehensive comparative study on the structural, electronic, and electrochemical properties of \NbC{2} and \NbCO{2} as anode materials for LIBs and SIBs is still lacking. A systematic theoretical investigation can provide valuable insights into their electrochemical behavior, guiding experimental efforts to optimize their performance for next-generation energy storage applications.

In this study, we employ first-principles density functional theory (DFT) calculations to investigate the fundamental electrochemical properties of \NbC{2} and \NbCO{2} MXenes as potential anodes for LIBs and SIBs. We systematically explore the adsorption energetics, ion diffusion barriers, charge transfer characteristics, and open-circuit voltage (OCV) profiles to assess their suitability for energy storage applications. Our findings reveal that \NbC{2} exhibits lower diffusion barriers, particularly for Na ions, making it a promising candidate for fast-charging SIBs. Conversely, \NbCO{2} demonstrates stronger adsorption and enhanced charge retention due to the presence of oxygen functional groups, albeit at the expense of increased ion migration resistance. Electronic structure analysis confirms the metallic nature of both MXenes, ensuring efficient electron transport crucial for battery operation.

By evaluating the trade-offs between adsorption strength, ion mobility, and charge retention, this study provides valuable insights into the potential of \NbC{2} and \NbCO{2} as anode materials for next-generation high performing LIBs and SIBs. The results not only highlight the advantages of sodium-ion batteries as a cost-effective alternative to lithium-based systems but also underscore the importance of surface functionalization in optimizing electrochemical performance. Our findings serve as a foundation for future research aimed at enhancing MXene-based anode materials through surface engineering and structural modifications to achieve high-efficiency energy storage.

\begin{figure}
  \centering
    \includegraphics[height=5.5cm]{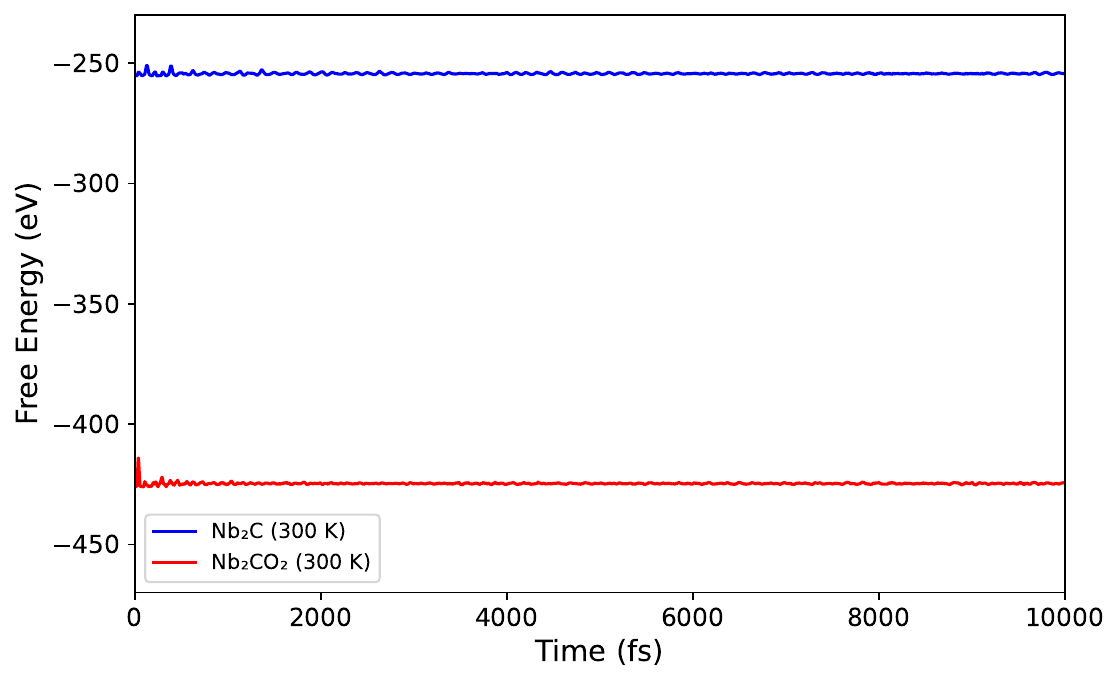}
    \caption{Free energy of a) \NbC{2} and b) \NbCO{2} through AIMD simulations for a duration of 10 ps at 300 K}
    \label{fgr:AIMD_sim}
\end{figure}

\section{Computational Method}
\textit{Ab initio} density functional theory (DFT) calculations in this study were performed using the Vienna Ab-initio Simulation Package (VASP)\cite{kresseEfficientIterativeSchemes1996}. The projector augmented wave (PAW) method was employed with a 500 eV energy cutoff to describe the electron properties of the system\cite{blochlProjectorAugmentedwaveMethod1994}. We constructed a 3 $\times$ 3 supercell with a 15 \text{\r{A}} vacuum layer for each structural model to eliminate interactions between periodic images. The Monkhorst pack scheme\cite{monkhorstSpecialPointsBrillouinzone1976} with 3 $\times$ 3 $\times$ 1 mesh points in k-space were used for all the systems considered in the present study. We used the Perdew-Burke-Ernzerhof (PBE) functional within the generalized gradient approximation (GGA) to optimize the geometric structures for the exchange-correlation potential\cite{perdewGeneralizedGradientApproximation1996}.

Geometrical relaxations were performed until the forces on the atoms are below 0.02 eV/\text{\r{A}} and the total energy difference was less than $1 \times 10^{-6}$ eV/atom. Van der Waals interactions were accounted for using the zero-damping DFT-D3 method of Grimme\cite{grimmeConsistentAccurateInitio2010}. 

Charge transfer during adsorption was analyzed by calculating the charge density difference (CDD), with the amount of charge transfer quantified using Bader charge analysis\cite{sanvilleImprovedGridbasedAlgorithm2007}. The climbing-image nudged elastic band (CI-NEB) method\cite{henkelmanClimbingImageNudged2000} implemented in VASP was used to determine transition states and minimum energy pathways or barriers associated with the diffusion of metal ions. To compute the off resonance Raman response, we calculated the derivative of the polarizability with respect to each normal mode for all the normal modes in the given systems. Phonons at the $\Gamma$-point were computed in VASP using the finite displacement method by setting IBRION = 5. Python script\cite{RamanscVASP2025} was used along with VASP calculations to obtain the derivative of the polarizability (or macroscopic dielectric tensor) with respect to that normal mode coordinate.

\begin{figure*}
  \centering
  \begin{subfigure}{0.48\textwidth}
    \centering
    \caption*{\raisebox{1.5\height}{\hspace{0mm}\textbf{\NbC{2}}}} % Adjust position
    \vspace{-3.5mm}
    \includegraphics[width=\textwidth]{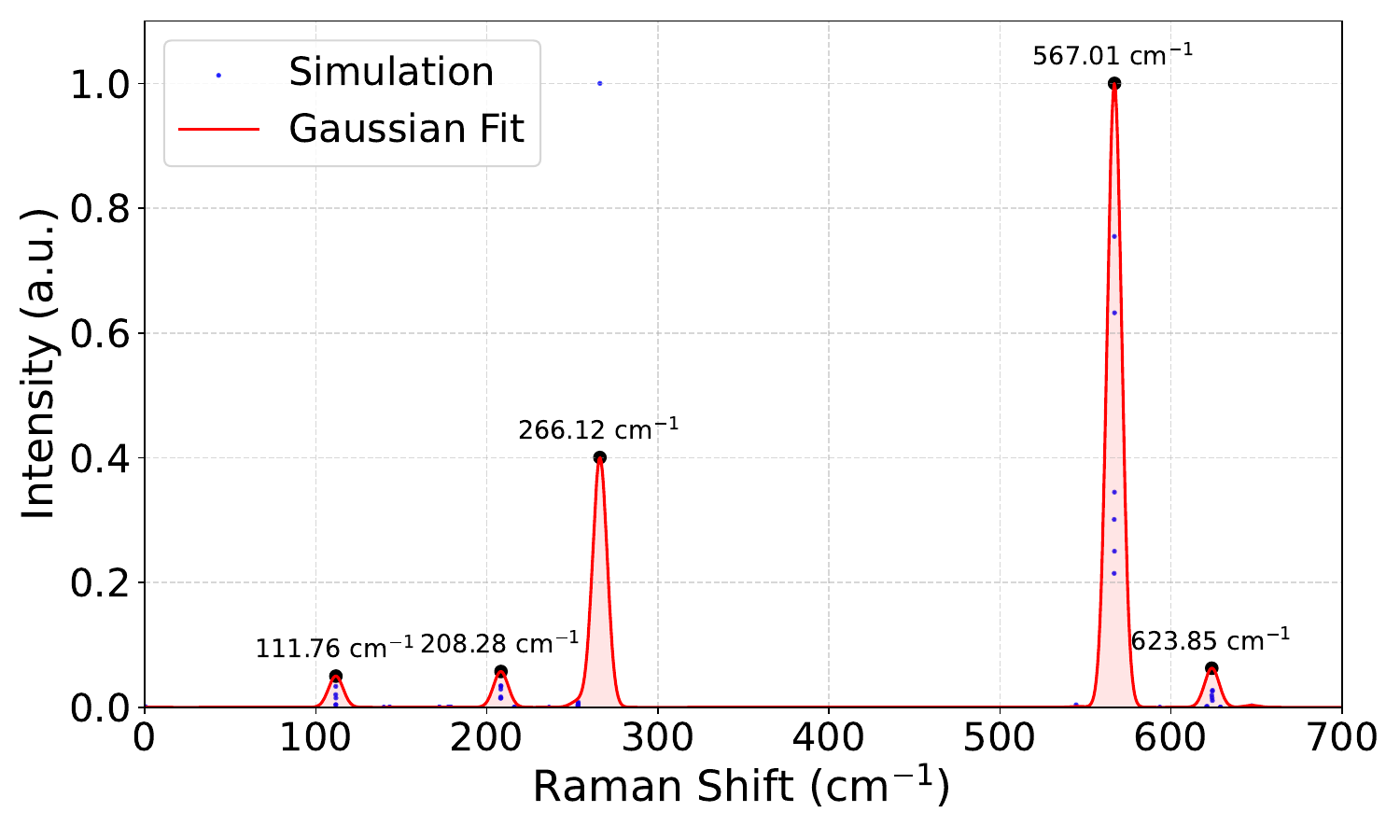}
  \end{subfigure}
  \begin{subfigure}{0.48\textwidth}
    \centering
    \caption*{\raisebox{1.5\height}{\hspace{0mm}\textbf{\NbCLi{2}}}} % Adjust position
    \vspace{-3.5mm}
    \includegraphics[width=\textwidth]{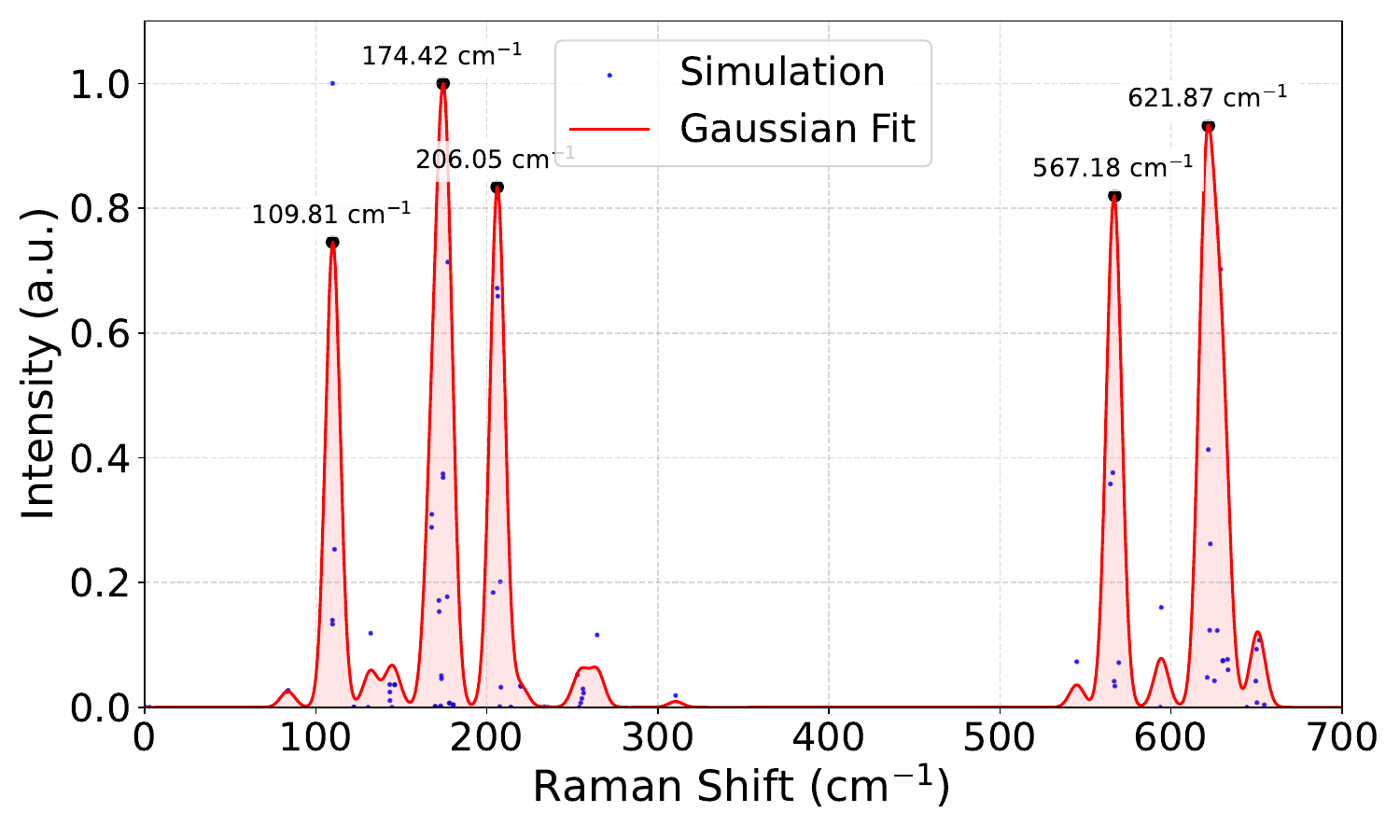}
  \end{subfigure}

  \vspace{-3mm} % Adjust vertical spacing

  \begin{subfigure}{0.48\textwidth}
    \centering
    \caption*{\raisebox{1.5\height}{\hspace{0mm}\textbf{\NbCO{2}}}} % Adjust position
    \vspace{-3.5mm}
    \includegraphics[width=\textwidth]{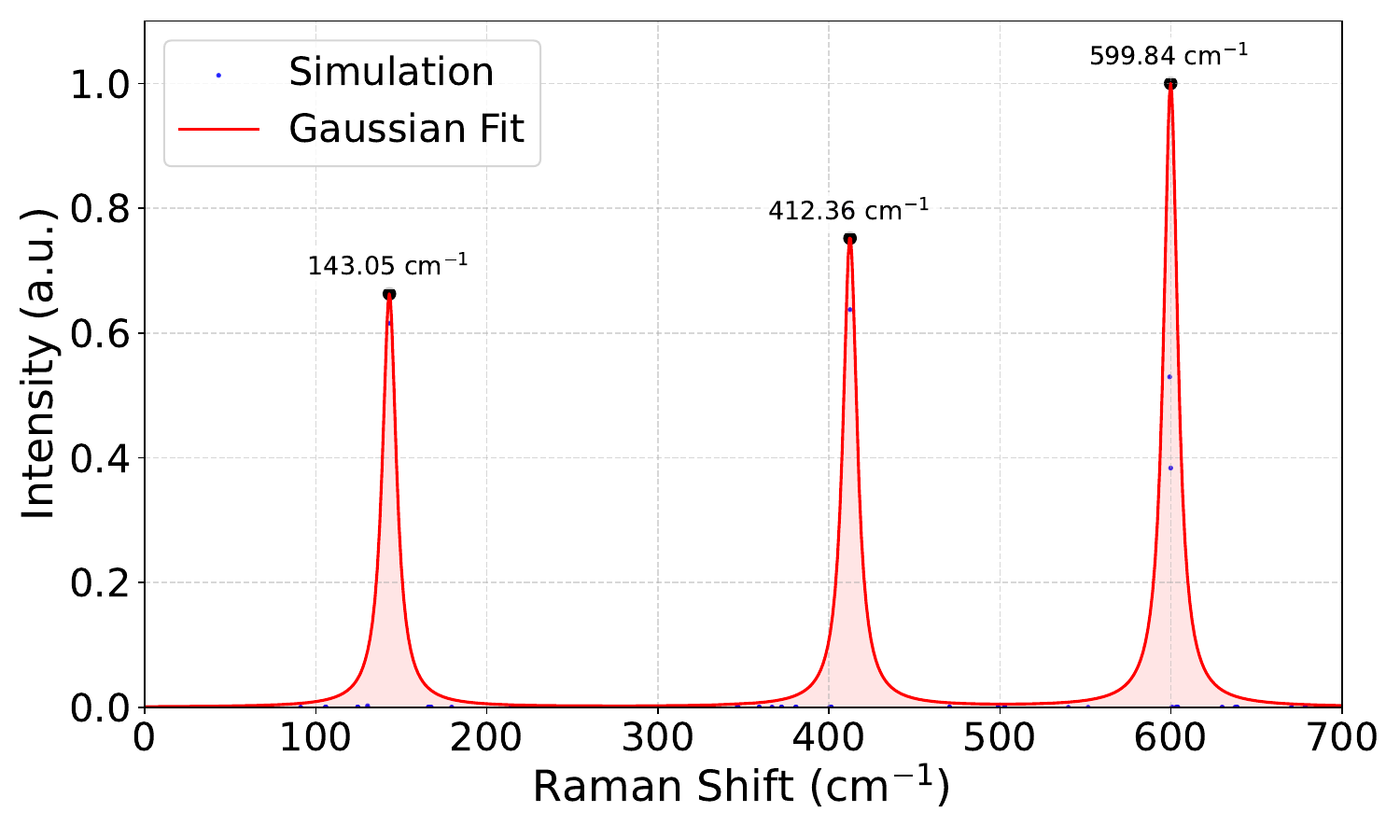}
  \end{subfigure}
  \begin{subfigure}{0.48\textwidth}
    \centering
    \caption*{\raisebox{1.5\height}{\hspace{0mm}\textbf{\NbCNa{2}}}} % Adjust position
    \vspace{-3.5mm}
    \includegraphics[width=\textwidth]{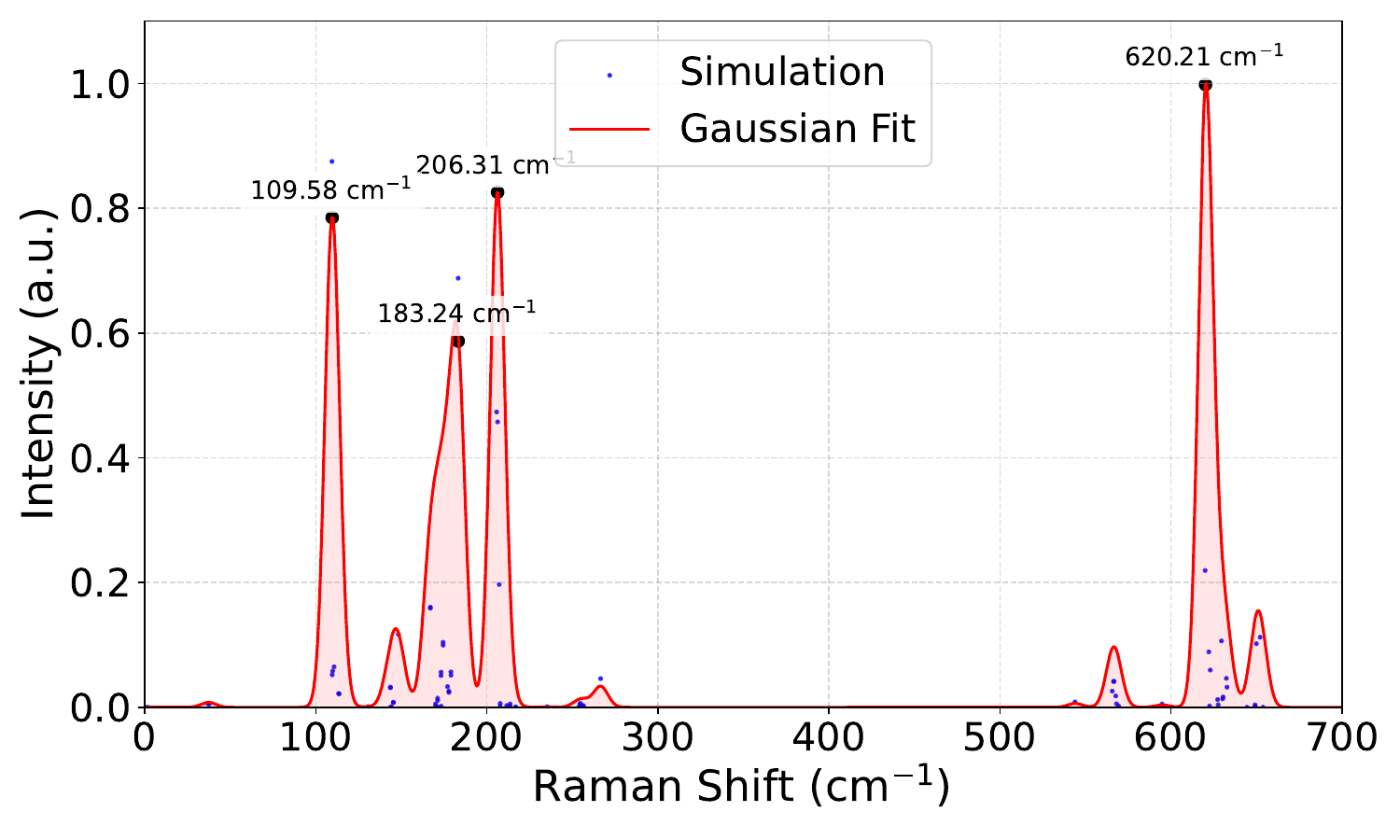}
  \end{subfigure}
  \caption{Raman Spectra of the \NbC{2}, \NbCO{2}, \NbCLi{2}, \NbCNa{2}.}
  \label{fgr:raman_Nb2Cs}
\end{figure*}

\begin{figure}
  \centering
  \includegraphics[width=\columnwidth]{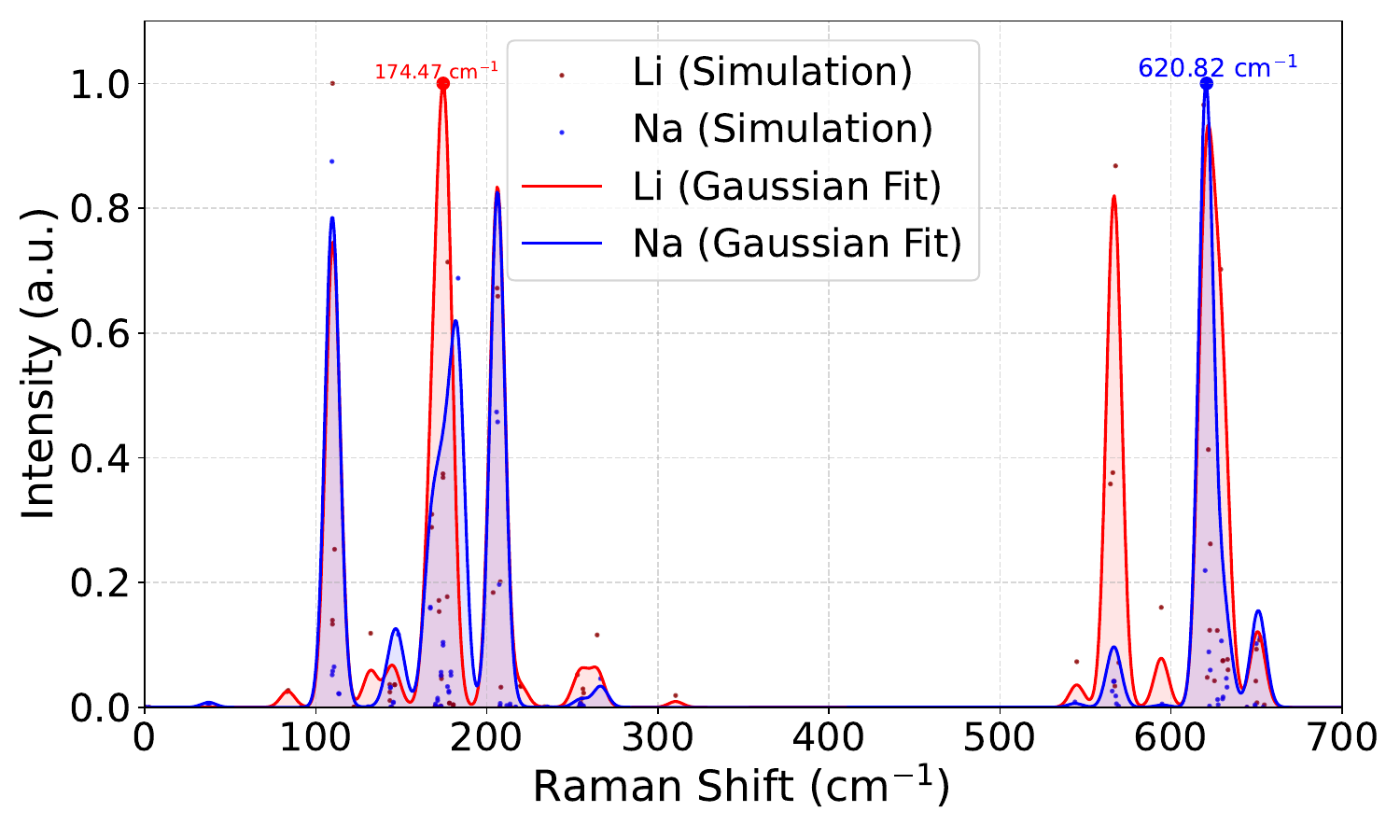}
  \caption{Comparison of Simulated Raman Spectra of \NbC{2} with Li and Na Intercalation: Impact on Peak Intensities and Electronic Environment.}
  \label{fgr:raman_sim_Nb2C-Na-Li}
 \end{figure}

 \section{Results}

 \subsection{Structural Properties}
 At the beginning of the present study, we fully relaxed the 2D crystal structure of \NbC{2} and \NbCO{2} Mxenes. The fully relaxed structure of \NbC{2} monolayer (Fig. \ref{fgr:atom_model_Nb2C_Nb2CO2}) exhibits a sandwich-like configuration consisting of three atomic layers arranged in a sequence of Nb-C-Nb. The optimized lattice parameter for bare \NbC{2} is a=3.14 \text{\AA} with a bond distance of 2.17 \text{\AA} between Nb-C and an interatomic distance of 3 \text{\AA} between Nb-Nb. We have considered two possible configurations (Fig. S1) for placing O atoms, one on top of C (Configuration I) and one on top of Nb (Configuration II). The energy comparison (Table S2) between the two examined configurations of oxygen indicates that it preferentially adopts the second arrangement, which has the lowest energy and is identified as the most stable state. The obtained cell parameter of  \NbCO{2} is 3.15 \text{\AA} with a Nb-O bond length of 2.12 \text{\AA}. The calculated bonding distance between Nb-C in \NbCO{2} is 2.22 \text{\AA}. Table \ref{tbl:lattice_parameters} outlines the structural properties of both MXenes.

 \begin{table}[h]
  \small
  \centering
  \caption{\ Lattice Parameters (a), Thickness (c), and Bond Length (d) of \NbC{2} and \NbCO{2}}
  \label{tbl:lattice_parameters}
  \begin{tabular*}{0.6\textwidth}{@{\extracolsep{\fill}}|l c c c c|}
    \hline
    System & \makecell{Lattice \\ Parameters (a)} & \makecell{Thickness \\ (c) \text{\AA}} & \makecell{Bonded \\ Atoms} & \makecell{Bond Length \\ (d) \text{\AA}}\\
    \hline
    \multirow{1}{*}{\NbC{2}} & 3.14 & 2.98 & Nb-C  & 2.17\\
    \hline
    \multirow{2}{*}{\NbCO{2}} & \multirow{2}{*}{3.15} & \multirow{2}{*}{4.78} & Nb-C  & 2.22\\
                              & & & Nb-O & 2.12\\
    \hline
  \end{tabular*}
\end{table}

These parameters are consistent with previous experimental values\cite{zhangRegulationCOOxidation2021,houNb2CO2PromisingSensor2024}. Notably, For \NbCO{2}, the Nb-C bond length is extended, suggesting that oxygen groups interact strongly with the bare \NbC{2} monolayer. Fig. \ref{fgr:AIMD_sim} presents the molecular Dynamics simulations conducted at 300 K over a 10 ps timescale of \NbC{2} and \NbCO{2} structures. %\todo{the meaning of this is not clear to me. Could you be more specific? response: Over the 10 ps time period it doesn't show much variation in the free energy, thus the assumption is that the strucutre is exhibit thermodynamic stability.}%
 Both structures showed thermodynamic stability throughout the observed period. \NbCO{2} maintained a lower energy than \NbC{2}, indicating greater structural stability over time. %\todo{Is \NbCO{2} a 2D structure? response: Yes, \NbCO{2} is a trigonal 2D structure where a=b!=c but since this a 2D structure, we don't need to report c.}%

\subsection{Raman Spectroscopy Analysis of \NbC{2} and \NbCO{2} with Li/Na Intercalation}

The phonon spectrum of a material is essential for understanding various macroscopic properties, including electrical and thermal conductivities. Previous studies have reported the phonon spectra of both \NbC{2} and \NbCO{2}, confirming their dynamic stability through the observation of non-negative frequencies\cite{yangElectronicStructuresElectron2020}. For pristine \NbC{2}, the three constituent atoms give rise to six optical modes and three acoustic modes. In contrast, for \NbCO{2}, the primitive cell contains five atoms, resulting in twelve optical modes and three acoustic modes. Among them, $E_g$ and $A_{1}g$ modes are found to be Raman active\cite{giordanoRamanSpectroscopyLaserInduced2024}.

To examine structural transformations, we computationally simulated the Raman spectra of \NbC{2} with and without Li/Na intercalation, as well as \NbCO{2}. The results of these simulations are shown in Fig. \ref{fgr:raman_Nb2Cs}. In case  of \NbC{2}, The peak at 208 $cm^{-1}$ is attributed to the $E_{1}g$ symmetry mode, representing an in-plane shear vibration involving Nb and C atoms experimentally obtained at 206 $cm^{-1}$\cite{linTwoDimensionalBiodegradableNiobium2017}. The Raman peak observed at 266.12 $cm^{-1}$ corresponds closely to the experimentally obtained A1g mode at 268 $cm^{-1}$\cite{huScreeningSurfaceStructure2018}, which arises from the out-of-plane vibrations of Nb and C atoms. We obtained another intense peak at  567.01 $cm^{-1}$ in the \NbC{2} structure which is present in both Li and Na intercalated \NbC{2} structure. The presence of various terminations and structural imperfections in the MXene lattice makes analyzing the Raman spectra complex\cite{adomaviciute-grabusoveMonitoringTi3C22024}. Although experimental validation is lacking for the obtained 567.01 $cm^{-1}$, our theoretical prediction of this peak may assist future investigations on the structure. 

For \NbCO{2}, a Raman peak was observed at 599.84 $cm^{-1}$, corresponding to the out-of-plane oxygen vibration ($A_{1}g$ mode). This value is in between the previously reported\cite{houNb2CO2PromisingSensor2024} theoretical prediction of 546 $cm^{-1}$ and the experimental measurement of 681 $cm^{-1}$. Another detected peak at 143.05 $cm^{-1}$ is attributed to the $E_g$ mode of \NbCO{2}, which was previously calculated at 132 $cm^{-1}$\cite{huScreeningSurfaceStructure2018}. Additionally, the peak at 174.47 $cm^{-1}$ in the \NbCLi{2} and \NbCNa{2} is associated with the $E_g$ mode of \NbC{2}, showing a slight deviation from the previously calculated value\cite{huScreeningSurfaceStructure2018} of 166.11 $cm^{-1}$.

Interestingly, after Li/Na intercalation, the Raman peak positions remain unchanged, while their intensities show variation, as shown in Fig. \ref{fgr:raman_sim_Nb2C-Na-Li}. According to the fundamental principles of Raman scattering, the peak position is primarily determined by vibrational frequencies, which depend on the bond stiffness and atomic masses. The absence of peak shifts suggests that Li/Na intercalation does not significantly alter the structural framework or bonding forces within \NbC{2}.

In contrast, the intensity of Raman peaks is influenced by the polarizability of the material, which determines how the electron cloud around atoms responds to incident light\cite{orlandoComprehensiveReviewRaman2021}. Li/Na atoms, being highly electropositive, donate electrons to the \NbC{2} substrate upon intercalation. This charge transfer modifies the local electronic environment, affecting the material's polarizability and consequently altering the Raman scattering cross-section. Additionally, intercalation can modify electron-phonon interactions, influencing how efficiently phonons couple to the incident light. Variations in intensity can also arise due to changes in the local dielectric environment, which can either enhance or suppress Raman signals.

Thus, the observed changes in Raman intensity without peak shifts indicate that Li/Na intercalation primarily affects the electronic properties of \NbC{2} rather than its structural integrity. This is consistent with the understanding that while peak position shifts are associated with structural distortions, intensity changes reflect modifications in charge distribution and electronic interactions\cite{houNb2CO2PromisingSensor2024}.

\subsection{Metal Adsorption and Diffusion}
To investigate the metal (Li, Na) intercalation mechanism, we examined three high-symmetry adsorption sites based on a previous study: Top, H3, and T4 on the \NbC{2} and \NbCO{2} surfaces\cite{santoy-floresNb2Nb2CO22024}, as illustrated in Fig. \ref{fgr:atom_model_Nb2C_Nb2CO2}. The most exposed atomic layer was used as a reference in both cases. The Top site corresponds to a Li atom positioned directly above an atom in the outermost layer. The T4 site places the Li atom above the second-most exposed layer, while the H3 site is aligned with the third-most exposed layer (cf. Fig. \ref{fgr:atom_model_Nb2C_Nb2CO2}). We calculated the adsorption energy by using the following equation:

\begin{equation}
  E_{ad}= \frac{{E_{\NbC{2}(O\textsubscript{2})}}_M-{E_{\NbC{2}(O\textsubscript{2})}}-x\mu_M}{x}
\end{equation}

\noindent where $E_{ad}$ is the adsorption energy, ${E_{\NbC{2}(O\textsubscript{2})}}_M$ is the total energy of the system with metal ions, $E_{\NbC{2}(O\textsubscript{2})}$ is the energy of the system without metals, $\mu_M$ represents the energy of the isolated metal, and x is the total number of inserted adatoms. Note that \NbC{2}(O\textsubscript{2}) represents either \NbC{2} or \NbCO{2} when appropriate. Table \ref{tbl:adsorption_parameters} represents the calculated $E_{ad}$ values of all the systems investigated in the present study. Negative values of the adsorbed energy in all the systems suggest favorable adsorption of both Li and Na onto \NbC{2} and \NbCO{2} Mxenes.

\begin{table}[h]
  \small
  \centering
  \caption{\ Calculated Energy Adsorption $E_{ads}$ (in eV) for Li and Na on \NbC{2} and \NbCO{2}}
  \label{tbl:adsorption_parameters}
  \begin{tabular*}{0.48\textwidth}{@{\extracolsep{\fill}}|l |c c c|}  % Use 'c' for centering
    \hline
     & \multicolumn{3}{c|}{$E_{ads}$ (eV)}  \\  
    %\hline
    & TOP & H3 & T4 \\
    \hline
    Li@\NbC{2} & -2.57 & -2.70 & -2.72\\
    %\hline
    Li@\NbCO{2} & -3.20 & -3.70 & -3.46\\
    %\hline
    Na@\NbC{2} & -2.38 & -2.44 & -2.45\\
    %\hline
    Na@\NbCO{2} & -3.01 & -3.43 & -3.34\\
    \hline
  \end{tabular*}
\end{table}

Our findings on Li intercalation in \NbC{2} and \NbCO{2} are consistent with previous studies\cite{santoy-floresNb2Nb2CO22024}, confirming that the T4 and H3 adsorption sites exhibit the highest stability, with adsorption energies of -2.72 eV and -3.70 eV, respectively. A similar trend was observed for Na intercalation, with adsorption energies of -2.45 eV and -3.43 eV.

\begin{figure}[h]
  \centering
    \includegraphics[width=\columnwidth]{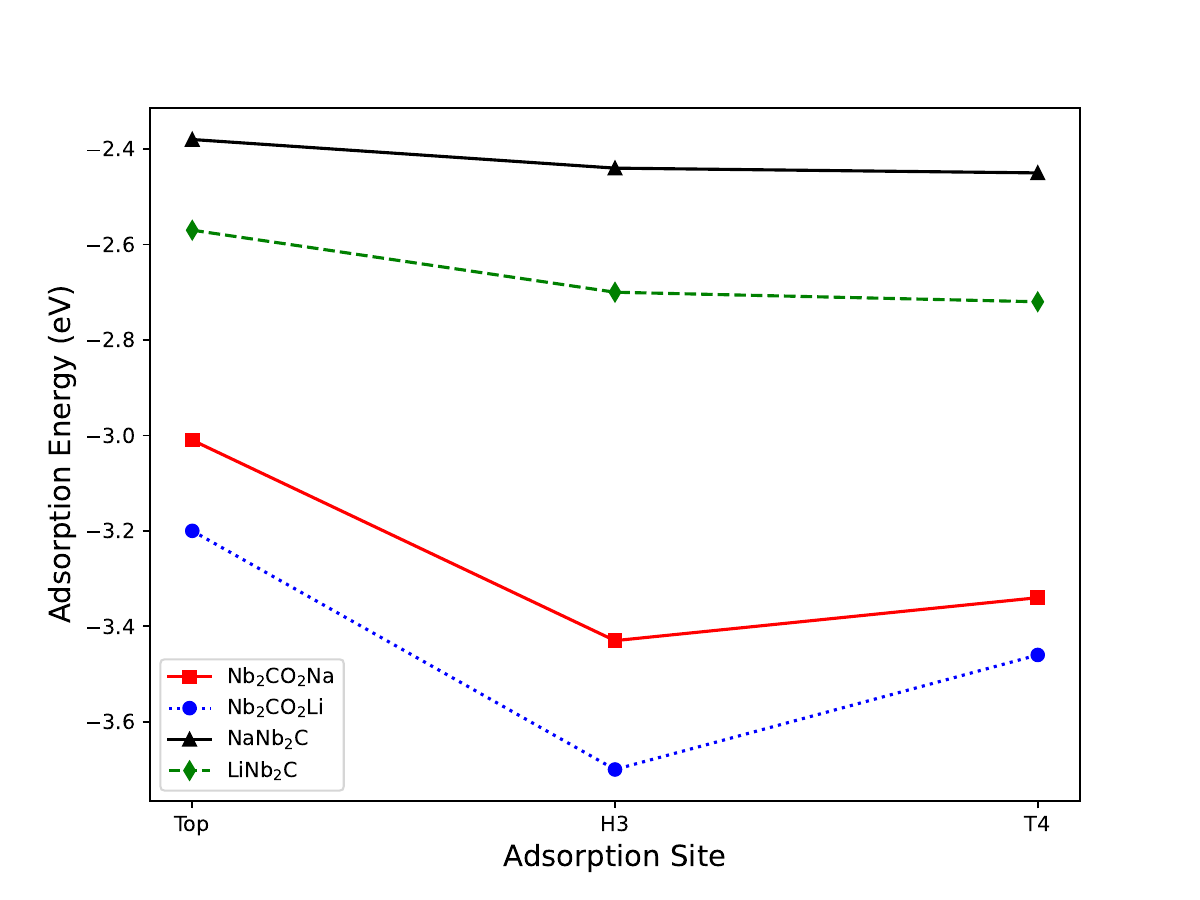}
    \caption{The line chart of adsorption energy of Li/Na on various sites of \NbC{2} and \NbCO{2} (eV).}
    \label{fgr:adsorption_energy}
\end{figure}

Fig. \ref{fgr:adsorption_energy} provides a comparative analysis of the adsorption energies. Notably, the energy difference between the H3 and T4 sites for Na is minimal, approximately 0.01 eV, suggesting that the H3 site in the Na-intercalated \NbC{2} system is also a viable adsorption site. However, in this study, we selected the T4 site for Na adsorption in the \NbC{2} system for all calculations, as it exhibited slightly lower energy than H3. Across all studied systems, \NbCO{2} demonstrated significantly lower adsorption energy compared to pristine \NbC{2}, indicating enhanced stability due to the presence of O functional groups.

\begin{figure*}
  \centering
  % First Column (a)
  \begin{minipage}{0.48\textwidth} % First column
    \centering
    \begin{overpic}[width=\textwidth]{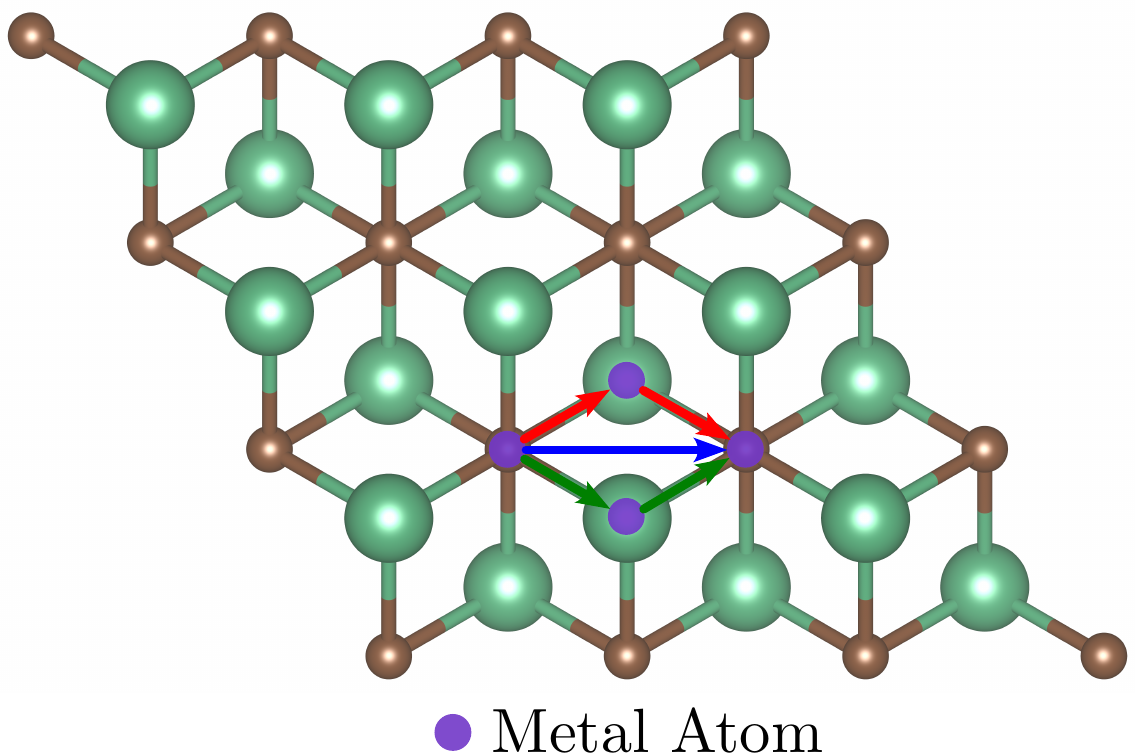}
      \put(85, 180){\textbf{(a)}} % Adjust (x, y) for positioning
    \end{overpic}
  \end{minipage}
  % Second Column (b) & (c)
  \begin{minipage}{0.48\textwidth} % Second column (stacked figures)
    \centering
    \begin{overpic}[width=\textwidth]{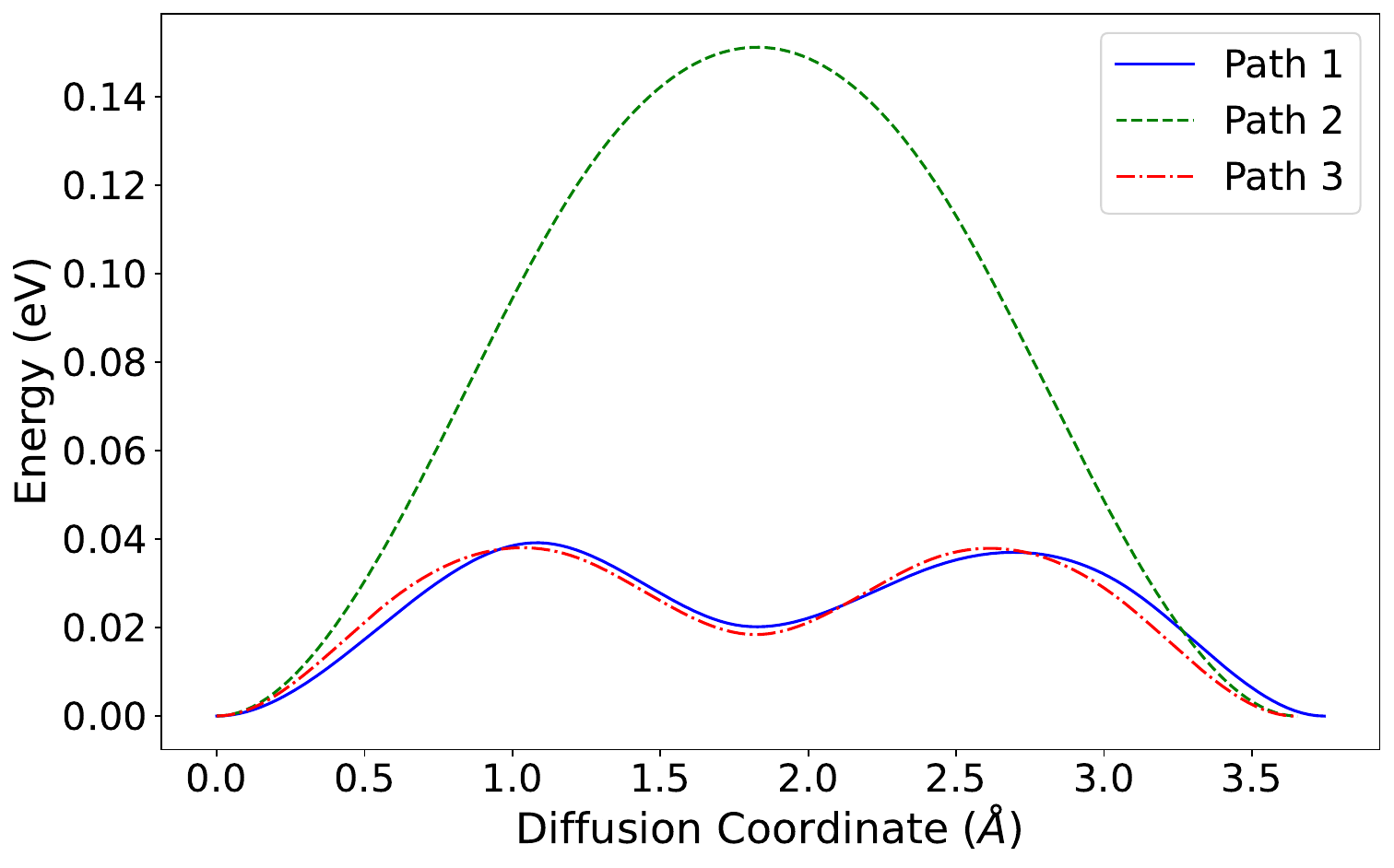}
      \put(35, 140){\textbf{(b)}} % Adjust position
    \end{overpic}

    \vspace{0mm} % Adjust spacing between (b) and (c)

    \begin{overpic}[width=\textwidth]{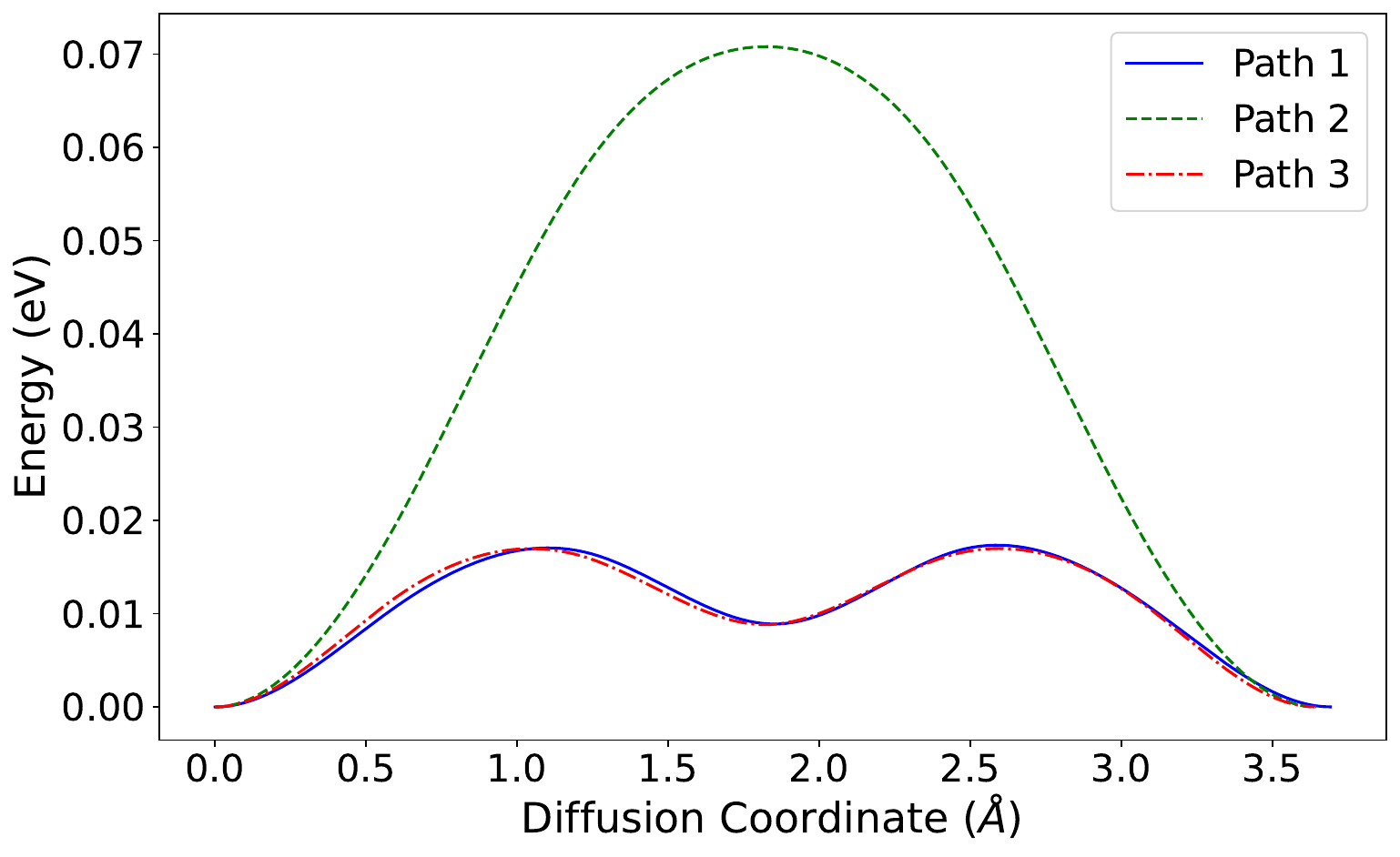}
      \put(35, 140){\textbf{(c)}} % Adjust position
    \end{overpic}
  \end{minipage}
  \vspace{-4mm}
  \caption{(a) Atomistic model showing the diffusion pathway for \NbC{2}. Path 1 (blue arrow), path 2 (green arrow), and path 3 (red arrow). (b) Diffusion barrier of Li atoms on \NbC{2} (c) Diffusion barrier of Na atoms on \NbCO{2}.}
  \label{fgr:metal_adsorption_site}
\end{figure*}

Efficient charge and discharge rates in rechargeable batteries rely on ion migration and electron transport kinetics. Lower diffusion barriers for Li-ion and Na-ion play a key role in enabling high-performance fast-charging and discharging. We employed the CI-NEB approach to gain insights on the Li-ion and Na-ion migration pathways and energy barriers in \NbC{2} and \NbCO{2} monolayers. As observed in Table \ref{tbl:adsorption_parameters}, the T4 site is identified as a favorable adsorption site for \NbC{2}, while the H3 site is preferred for \NbCO{2} during both Na and Li intercalation. Accordingly, in the CI-NEB simulations, the initial adsorption sites for Na and Li intercalation were set to T4 for \NbC{2} and H3 for \NbCO{2} to ensure accurate diffusion pathway analysis. We determined three potential diffusion pathways between the two nearest adsorption sites on the \NbC{2} and \NbCO{2} surface. Fig. \ref{fgr:metal_adsorption_site} shows the diffusion pathways on an atomistic model of \NbC{2}.%\todo[inline]{The font sizes for the tick mark labels needs to be increased. The x-label is partially covered up for (b).} 

Li/Na migration in \NbC{2} and \NbCO{2} follows distinct diffusion pathways influenced by the adsorption sites and atomic interactions within the monolayer. As Li/Na moves along Path 1 (straight pathway), migration occurs directly from the initial to the final adsorption site. In Path 2 (upper pathway), it traverses the upper-layer atoms of the monolayer. Meanwhile, along Path 3 (lower pathway), Li/Na passes through a second stable adsorption position on the surface. Computational results indicate that Path 2 exhibits the highest energy barrier for  \NbC{2} primarily due to the close proximity between Li/Na and the upper-layer atoms, restricting smooth migration.

\begin{figure}[h]
  \centering
    \includegraphics[width=\columnwidth]{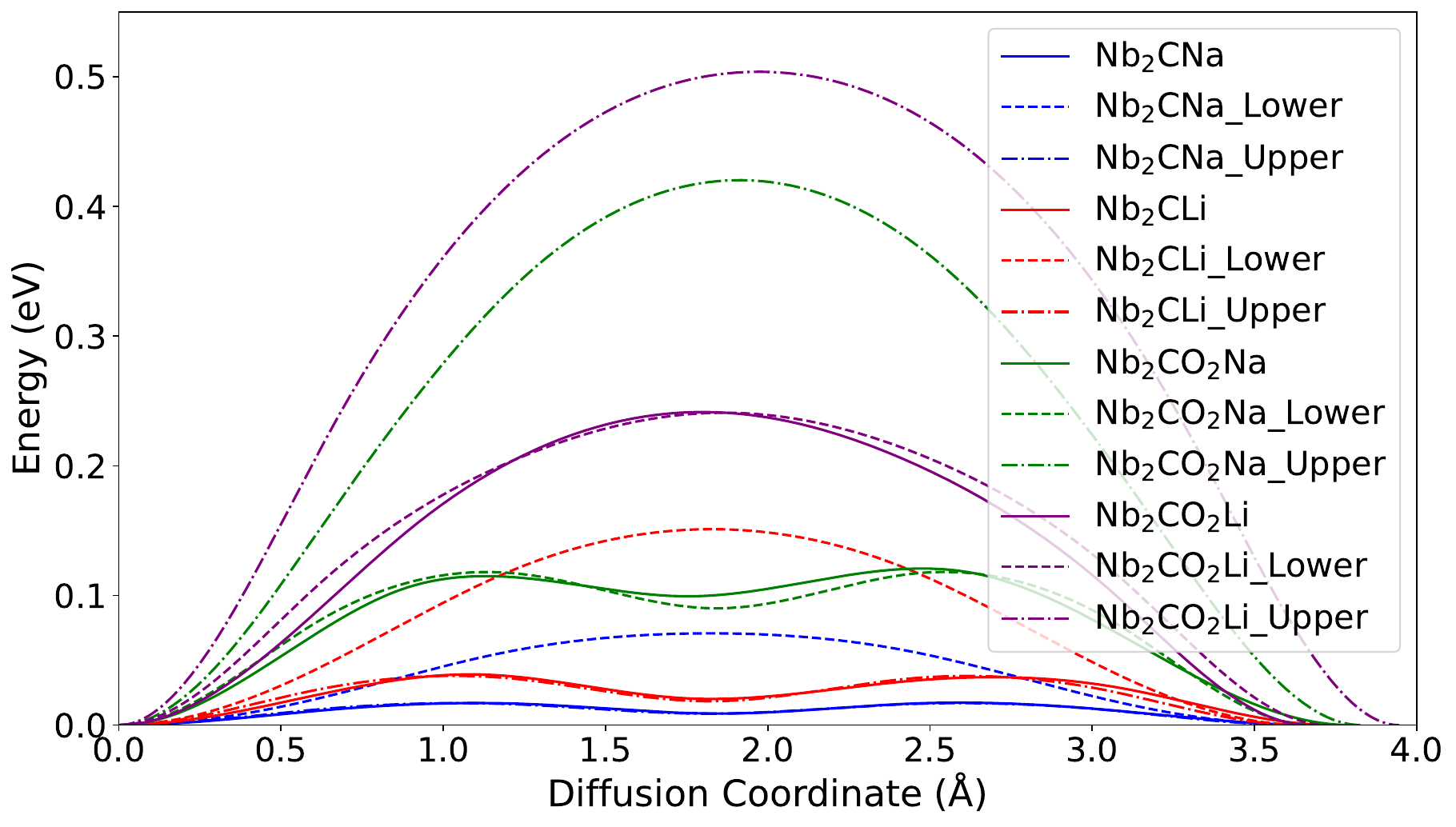}
    \caption{Diffusion paths and energy barriers for Na/Li atoms on  \NbCO{2}  and \NbCO{2}.}
    \label{fgr:diffusion_barrier_comprehensive}
\end{figure}

The calculated diffusion barriers indicate that \NbCO{2} exhibits higher energy barriers compared to bare \NbC{2} (Fig. \ref{fgr:diffusion_barrier_comprehensive}), %\todo[inline]{The plot title should be removed and all font sizes in this plot need to be increased}
primarily due to the strong interaction between the functional groups and migrating Li/Na atoms. However, Na-ion diffusion barriers are consistently lower than those of Li-ions on both surfaces, suggesting that \NbC{2} and \NbCO{2} could be more suitable for Na-ion battery applications. For \NbCNa{2}, the diffusion barriers for Path 1 (straight pathway), Path 2 (upper pathway), and Path 3 (lower pathway) are 16.90 meV, 16.21 meV, and 70.70 meV, respectively. Similarly, for \NbCLi{2}, the barriers for Path 1, Path 2, and Path 3 are 39.10 meV, 36.11 meV, and 151.10 meV, respectively. On the \NbCO{2} surface, Na-ion diffusion barriers are 117.07 meV for Path 1, 420.29 meV for Path 2, and 117.20 meV for Path 3. Meanwhile, Li-ion diffusion on \NbCO{2} shows the highest energy barriers, with values of 241.37 meV for Path 1, 441.64 meV for Path 2, and 240.92 meV for Path 3. Across all cases, Path 2 (upper pathway) consistently exhibits the highest energy barrier, particularly for \NbCO{2}, where the presence of functional groups significantly restricts ion mobility. Additionally, for Path 1 and Path 3 of Nb2C surface and path 3 of \NbCO{2} surface two peaks of nearly equal barrier values are observed, with a local minimum in between, corresponding to a metastable adsorption site for Li/Na atoms. This intermediate position allows temporary stabilization before continuing migration, influencing the overall diffusion process. Given these findings, \NbC{2} and \NbCO{2} surfaces provide a more efficient migration pathway for Na-ions compared to Li-ions, making them highly suitable for Na-ion battery applications due to their lower diffusion resistance and smoother ion transport.

\begin{figure*}
  \centering
  \begin{subfigure}{0.22\textwidth}
    \centering
    \caption*{\raisebox{1.5\height}{\hspace{13mm}\textbf{(a) \NbCLi{2}}}} % Adjust position
    \vspace{-3mm}
    \includegraphics[width=\textwidth]{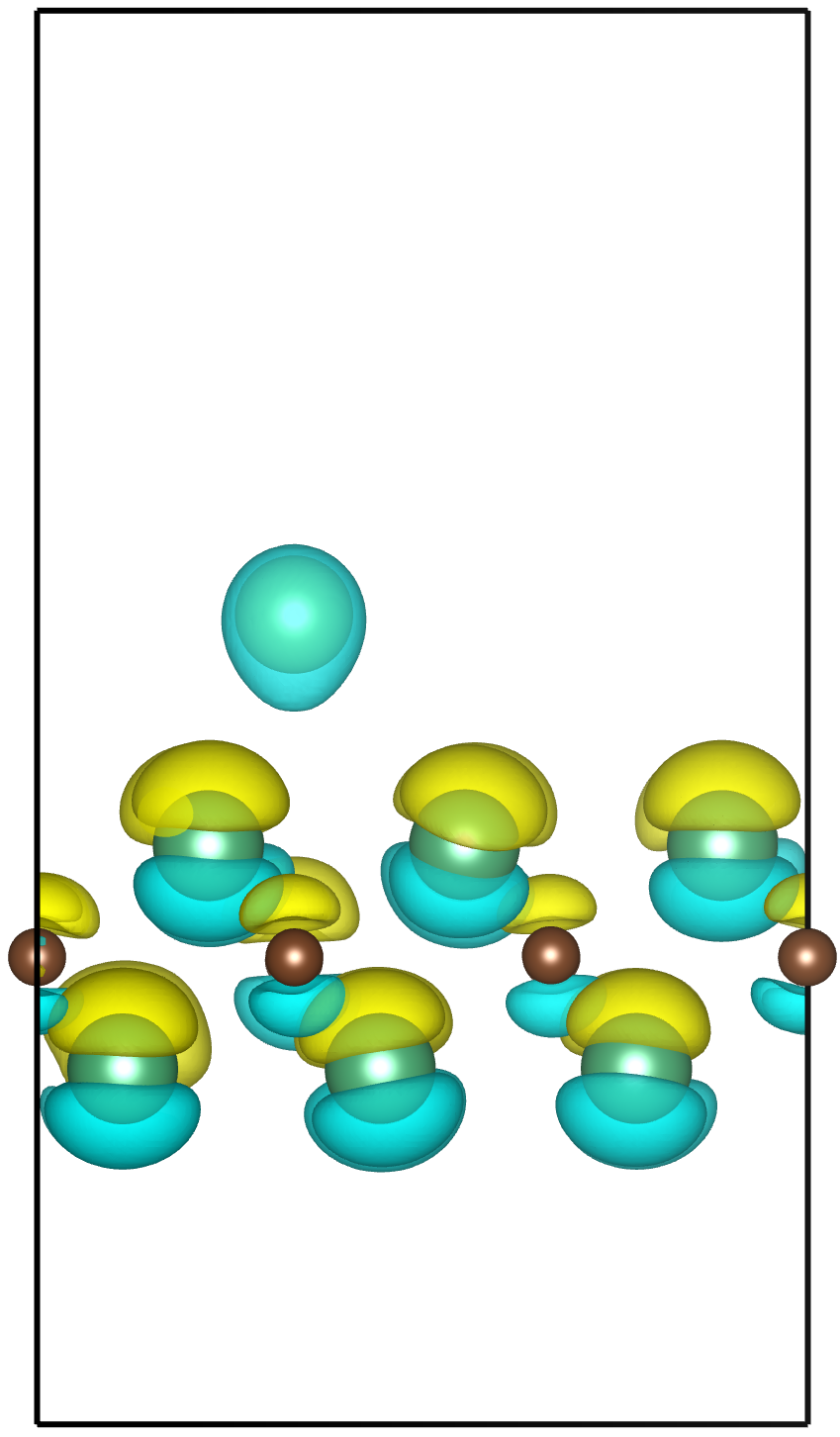}
    % \caption*{(a)}
  \end{subfigure}
  \begin{subfigure}{0.267\textwidth}
    \centering
    \caption*{\raisebox{1.5\height}{\hspace{13mm}\textbf{(b) \NbCO{2}Li}}} % Adjust position
    \vspace{-3mm}
    \includegraphics[width=\textwidth]{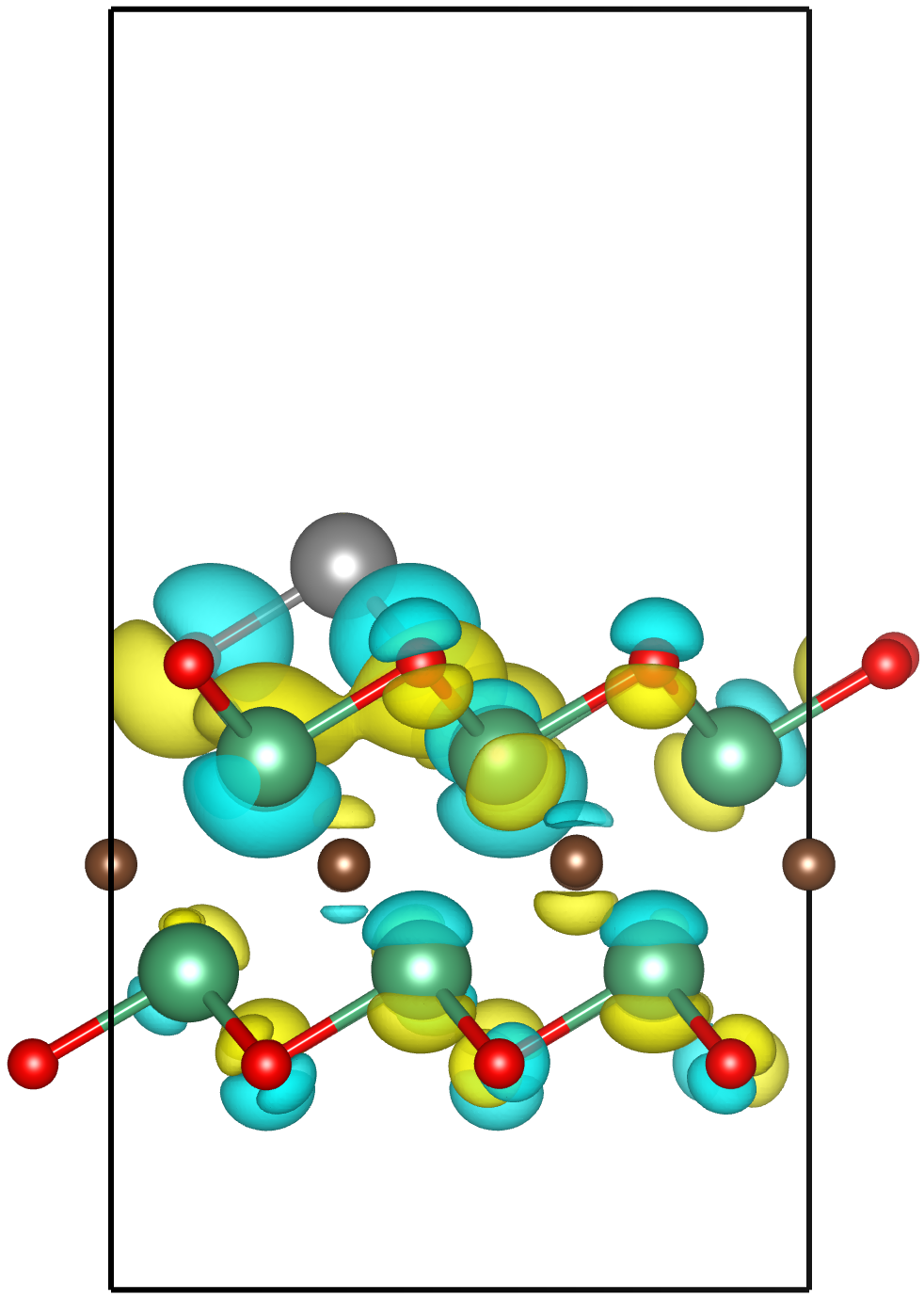}
  \end{subfigure}
  % \hspace{6mm}
  \begin{subfigure}{0.221\textwidth}
    \centering
    \caption*{\raisebox{1.5\height}{\hspace{13mm}\textbf{(c) \NbCNa{2}}}} % Adjust position
    \vspace{-3mm}
    \includegraphics[width=\textwidth]{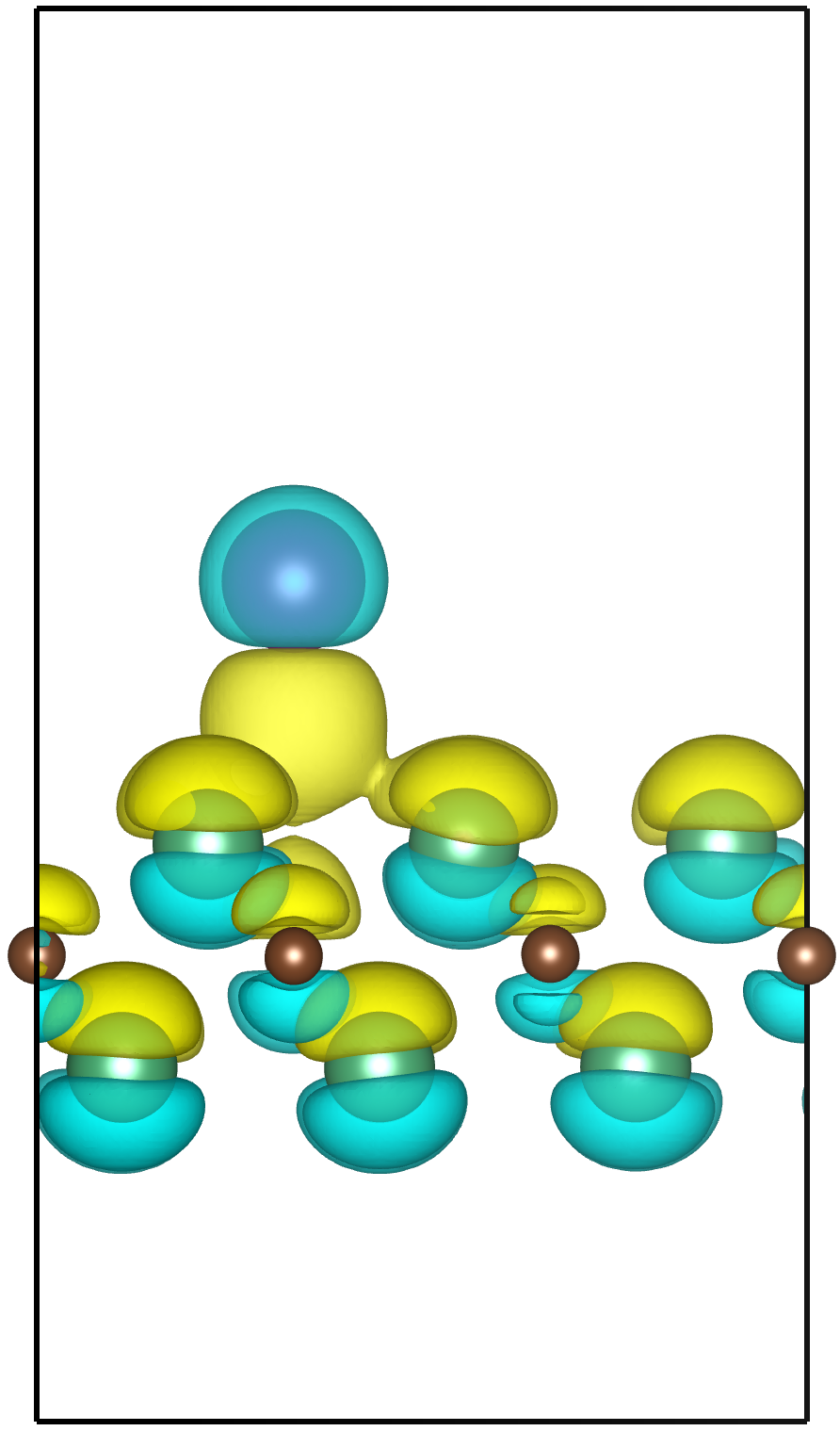}
  \end{subfigure}
  \begin{subfigure}{0.265\textwidth}
    \centering
    \caption*{\raisebox{1.5\height}{\hspace{13mm}\textbf{(d) \NbCO{2}Na}}} % Adjust position
    \vspace{-3mm}
    \includegraphics[width=\textwidth]{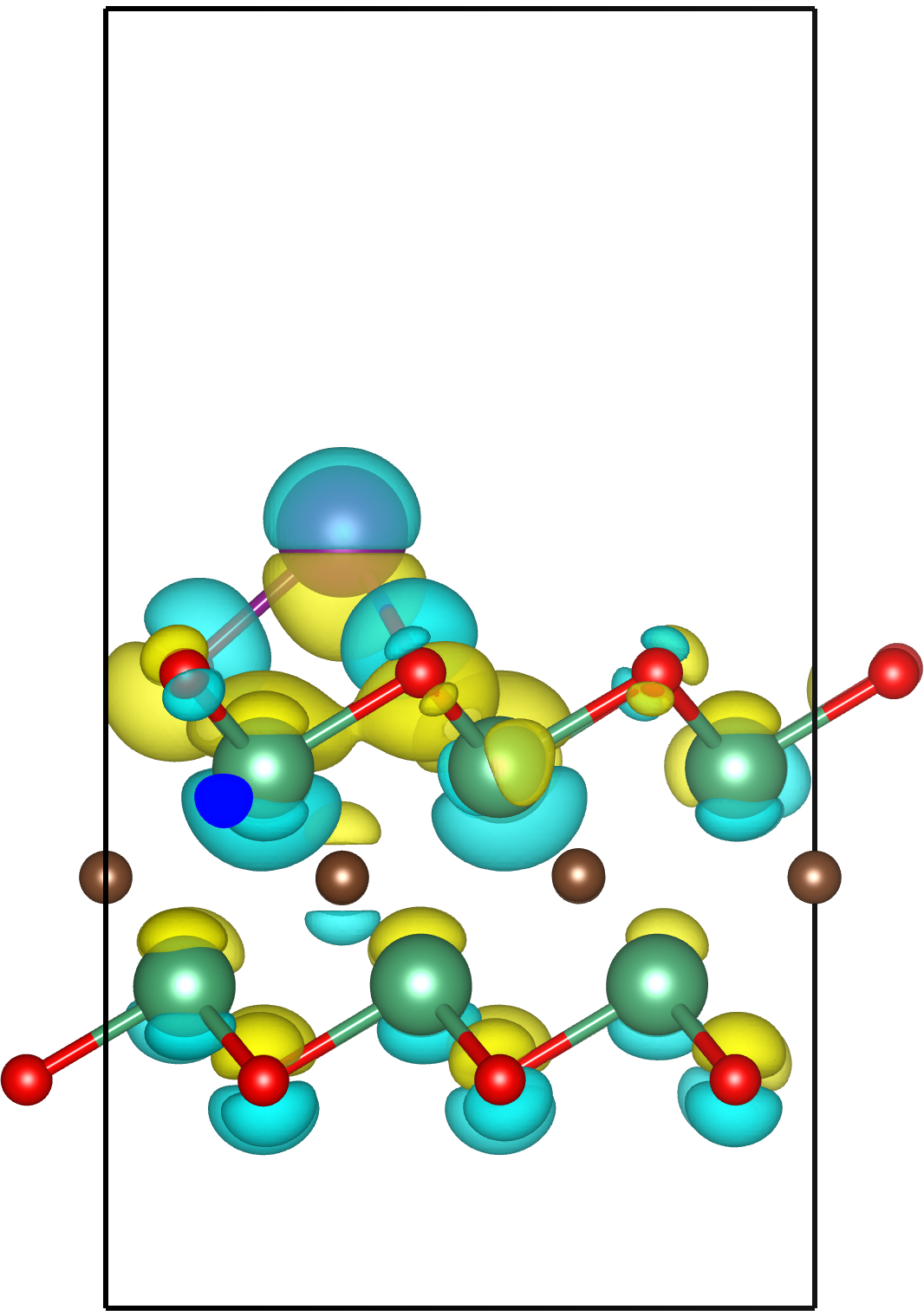}
  \end{subfigure}
  \vspace{-2mm}
  \caption{The charge density distribution  of a) \NbCLi{2} b) \NbCO{2}Li c) \NbCNa{2} and d) \NbCO{2}Na monolayer, where yellow and blue parts indicate accumulation and electron depletion (isosurface = 0.009 e/\text{\AA}).}
  \label{fgr:charge_density}
\end{figure*}

\subsection{Electronic Properties}

\subsubsection{Charge Transfer}

We plotted the charge difference density to further understand the charge transfer of these systems. Charge differences were calculated using the following equation:

\begin{equation}
  \Delta \rho = \rho_{\NbC{2}(O\textsubscript{2})_M} - \rho_{\NbC{2}(O\textsubscript{2})} - \rho_{M}
\end{equation}

In the equation, $\rho_{\NbC{2}(O\textsubscript{2})_M}$ is the charge density of \NbC{2}M or \NbCO{2}M with M representing either Na or Li, $\rho_{\NbC{2}(O\textsubscript{2})}$ is the charge density of \NbC{2} or \NbCO{2}, and $\rho_{M}$ is the charge density of Na or Li. The charge density distribution shows the  electron transfer from Na and Li to the substrate, signifying a chemical interaction between the metal ions and the surface. The CDD visualization (Fig. \ref{fgr:charge_density})
% \todo[inline]{This figure would be improved if subfigure (a) had Nb$_2$CLi written on it directly, (b) had Nb$_2$CO$_2$Li, etc}
reveals charge accumulation and depletion patterns across the \NbC{2} and \NbCO{2} surfaces. On both \NbC{2} and \NbCO{2} , a pronounced yellow region is observed around the Nb and O atom, indicating charge accumulation for both Li and Na adsorption. In contrast, the intercalated Li and Na atoms mostly  exhibits blue regions, reflecting the charge depletion. 
% \todo[inline]{I think something more specific than ``more extensively'' would strengthen the argument here}. 
The strong charge transfer observed at the interface aligns with Bader charge results, confirming the nearly complete ionization of Li and Na and their strong chemical adsorption onto the \NbCO{2} substrate.

\begin{table}[h]
  \small
  \centering
  \caption{\ Bader charge (in eV) of Li and Na on \NbC{2} and \NbCO{2}}
  \label{tbl:bader_charge}
  \begin{tabular*}{0.48\textwidth}{|@{\extracolsep{\fill}}l c|}  % Use 'c' for centering
    \hline
    System & \makecell{Bader Charge} \\  
    \hline
    \NbCLi{2} & 0.78 \\
    %\hline
    \NbCNa{2} & 0.37 \\
    %\hline
    \NbCO{2}Li & 0.91 \\
    %\hline
    \NbCO{2}Na & 0.91 \\
    \hline
  \end{tabular*}
\end{table}

The Bader charge analysis reveals distinct differences in charge transfer behavior between Li and Na ions on \NbC{2} and \NbCO{2} surfaces, highlighting the impact of oxygen functionalization on adsorption stability. On the \NbC{2} surface, Li transfers 0.78 e, indicating stronger electron donation compared to Na, which transfers only 0.37 e. This lower charge transfer suggests a weaker interaction for Na on \NbC{2}, making the adsorption less stable. However, oxygen-functionalized \NbCO{2} significantly enhances charge transfer, with both Li and Na donating 0.91 e. This improvement suggests that O-groups create stronger electrostatic interactions with the adsorbed ions, thereby increasing adsorption stability. The ability of \NbCO{2} to facilitate higher charge transfer makes it a more effective platform for ion retention, which could enhance battery performance by improving charge separation and reducing energy losses. This behavior is particularly important for Na-ion batteries, where efficient charge storage is crucial for long-term cycling stability.

\NbCO{2} provides stronger adsorption sites for Li and Na due to the higher charge transfer observed in Bader analysis. This stronger interaction is primarily due to the electronegative oxygen functional groups, which create more localized charge accumulation around the adsorbed ions. As a result, Li and Na ions experience greater electrostatic attraction, leading to higher stability on \NbCO{2} compared to \NbC{2}. However, this enhanced adsorption also contributes to higher diffusion barriers, as observed in the diffusion barrier analysis. Since the ions are more strongly bound to \NbCO{2}, they require more energy to overcome the adsorption potential and migrate across the surface. This is reflected in the higher energy barriers for both Li and Na on \NbCO{2}, particularly in Path 2 (upper pathway), where the interaction with the surface is most pronounced. In contrast, the weaker adsorption of Na on \NbC{2} results in lower charge transfer but also lower diffusion barriers, allowing for easier ion migration.

\begin{figure*}
  \centering
  \begin{subfigure}{0.48\textwidth}
    \centering
    \caption*{\raisebox{1.5\height}{\hspace{45mm}\textbf{(a) \NbC{2}}}} % Adjust position
    \vspace{-3.5mm}
    \includegraphics[width=\textwidth]{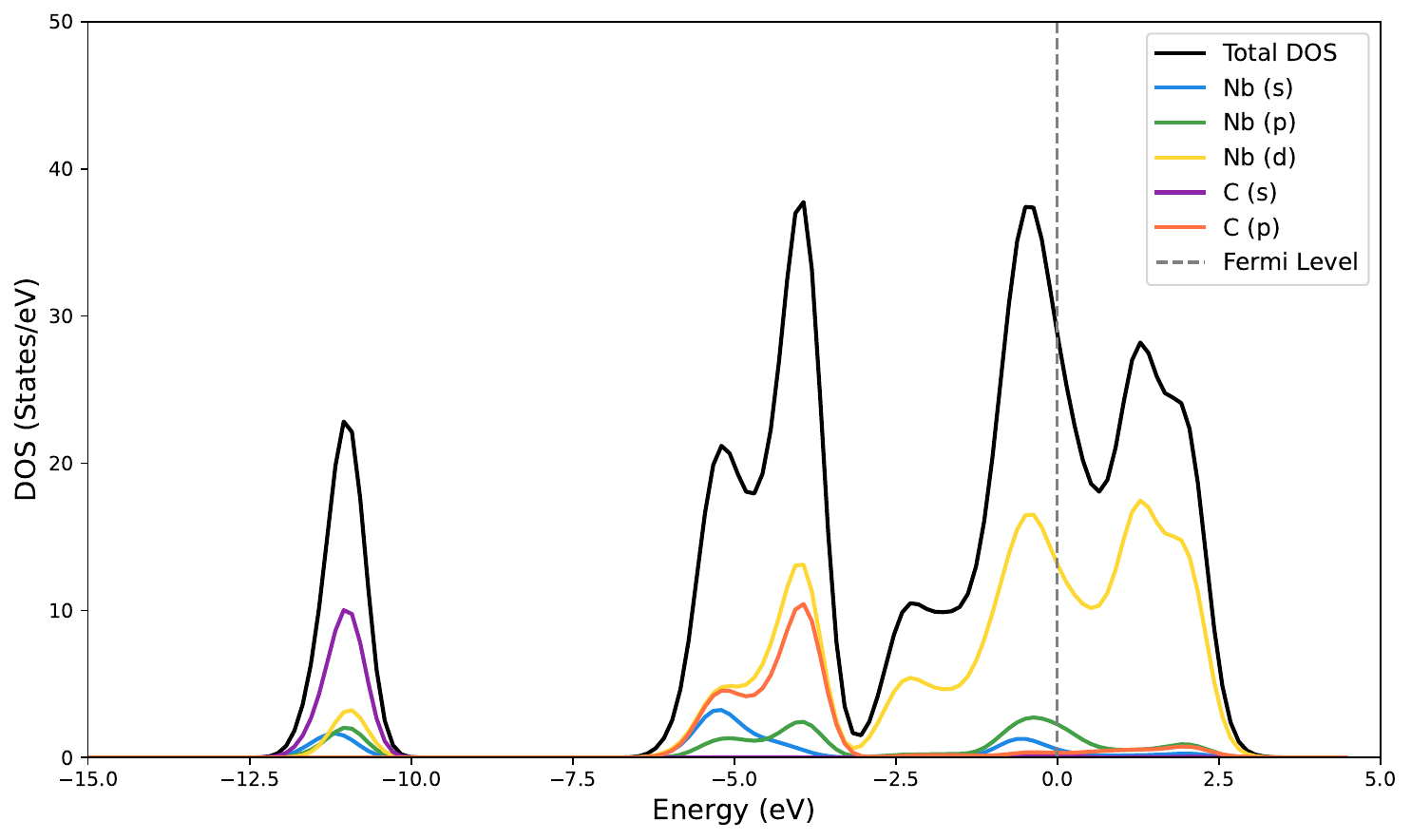}
  \end{subfigure}
  \begin{subfigure}{0.48\textwidth}
    \centering
    \caption*{\raisebox{1.5\height}{\hspace{45mm}\textbf{(b) \NbCO{2}}}} % Adjust position
    \vspace{-3.5mm}
    \includegraphics[width=\textwidth]{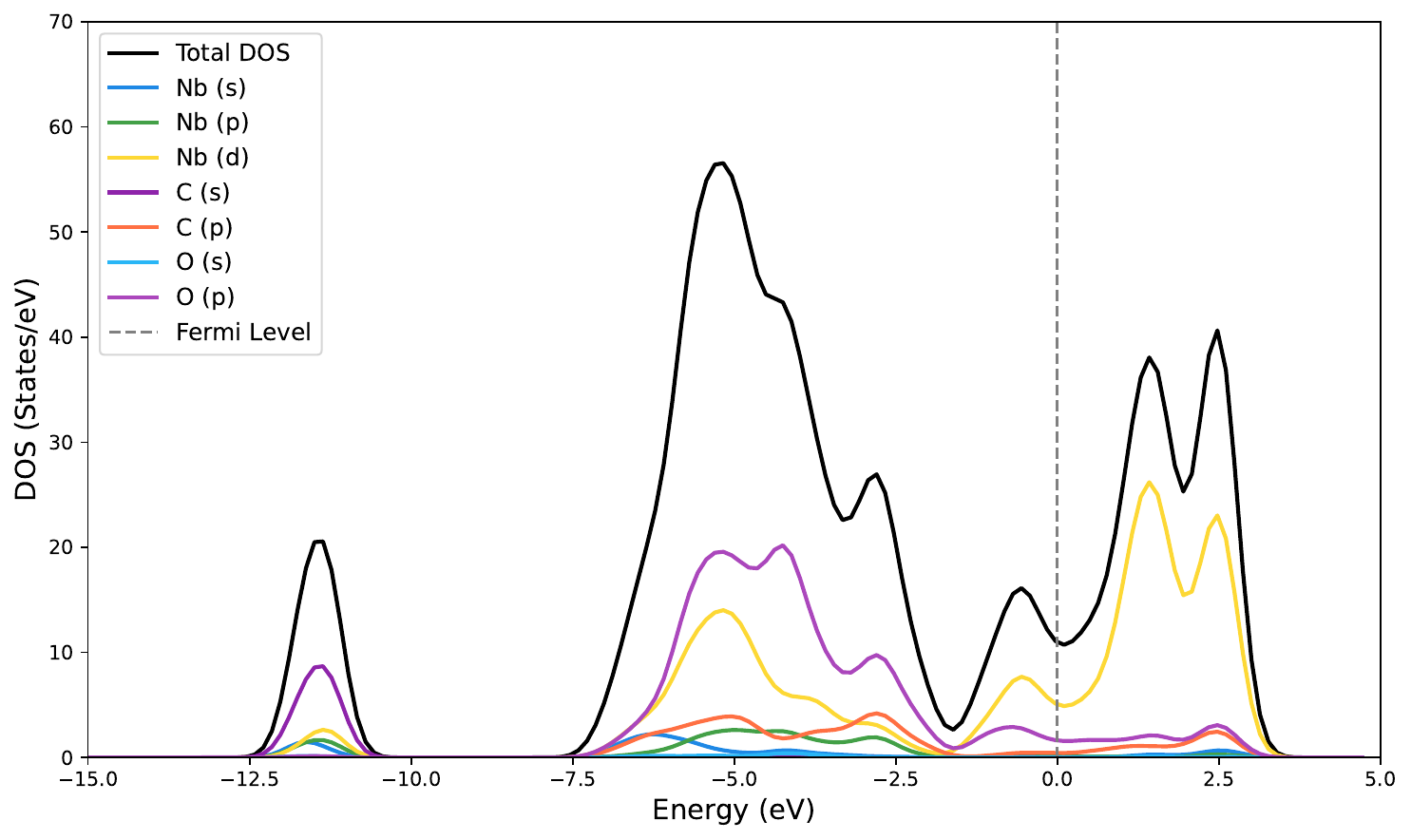}
  \end{subfigure}

  \caption{a) Electronic band structure along the $\Gamma-M-K-\Gamma$ path and their corresponding projected density of states for (a) \NbC{2} and (b) \NbCO{2} MXenes.}
  \label{fgr:electronic_band_structure}
\end{figure*}

These findings highlight a critical trade-off in anode material optimization: while strong adsorption enhances charge retention and stability, it can also hinder ion mobility, which is crucial for fast charge and discharge rates in high-performance batteries. The lower diffusion barriers of Na suggest that sodium-ion batteries could achieve more efficient ion transport compared to Li-ion batteries, making them a viable alternative for energy storage applications. However, the significantly higher barriers in \NbCO{2}, despite improved charge transfer, indicate the need to balance adsorption strength with diffusion kinetics to optimize anode performance. Tailoring the surface functionalization of \NbCO{2} or introducing structural modifications to reduce migration resistance could provide an effective strategy to enhance its role as a cost-effective anode material for both Na- and Li-ion batteries.

\subsubsection{Partial Density of States}

\begin{figure*}
  \centering
  \begin{subfigure}{0.48\textwidth}
    \centering
    \caption*{\raisebox{1.5\height}{\hspace{40mm}\textbf{a) \NbC{2}Na}}} % Adjust position
    \vspace{-3.5mm}
    \includegraphics[width=\textwidth]{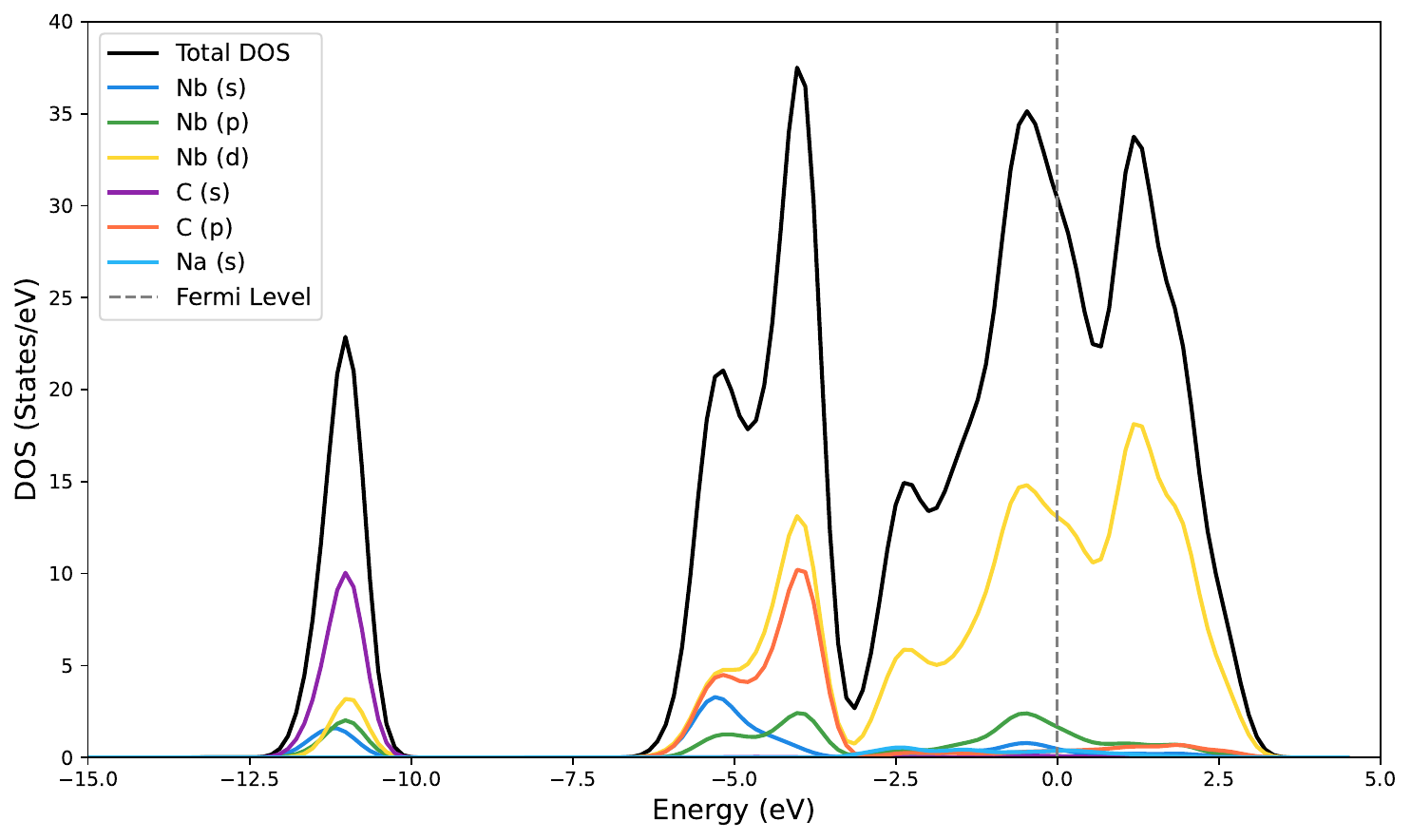}
  \end{subfigure}
  \begin{subfigure}{0.48\textwidth}
    \centering
    \caption*{\raisebox{1.5\height}{\hspace{40mm}\textbf{b) \NbCO{2}Na}}} % Adjust position
    \vspace{-3.5mm}
    \includegraphics[width=\textwidth]{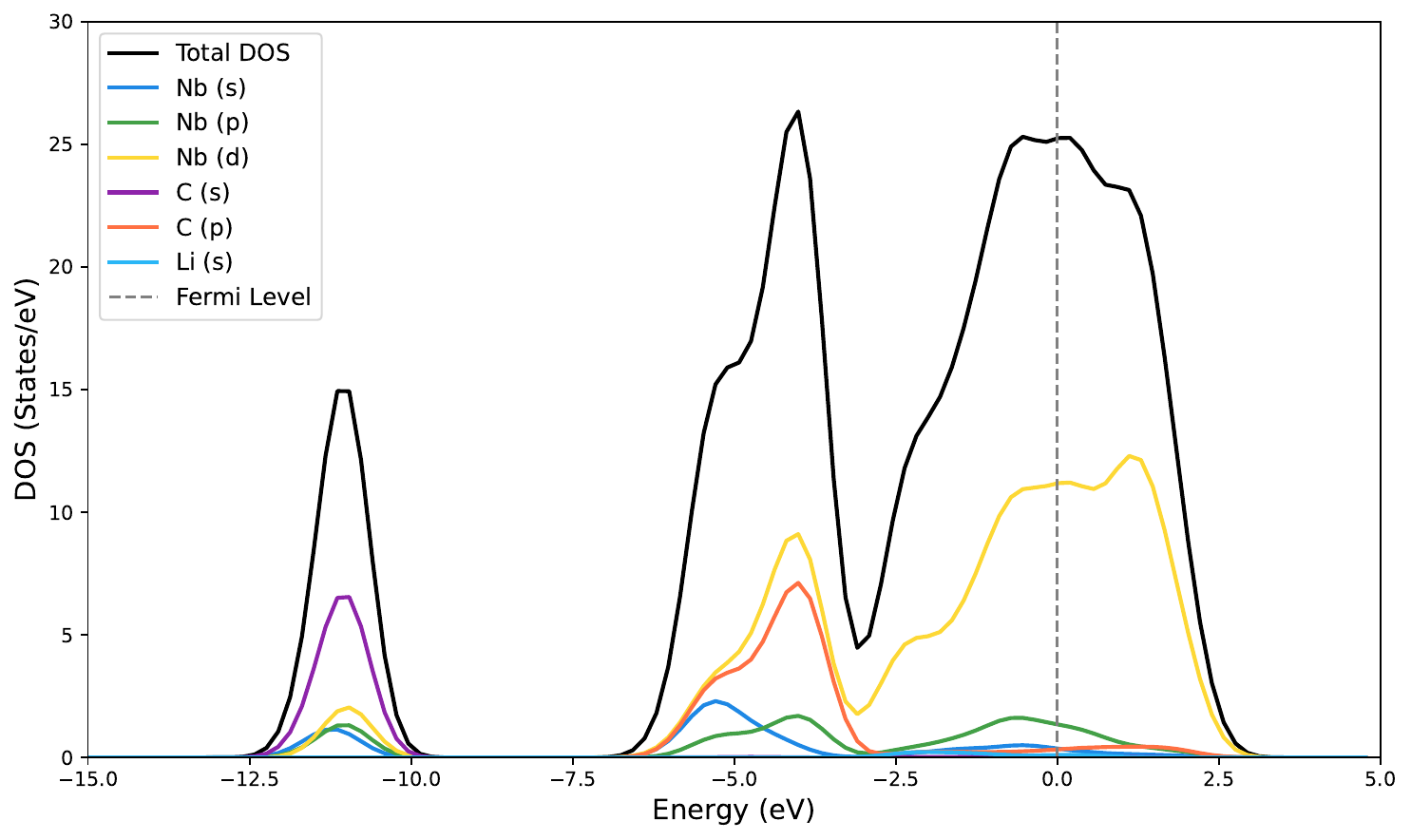}
  \end{subfigure}

  \vspace{-2mm} % Adjust vertical spacing

  \begin{subfigure}{0.48\textwidth}
    \centering
    \caption*{\raisebox{1.5\height}{\hspace{40mm}\textbf{c) \NbC{2}Li}}} % Adjust position
    \vspace{-3.75mm}
    \includegraphics[width=\textwidth]{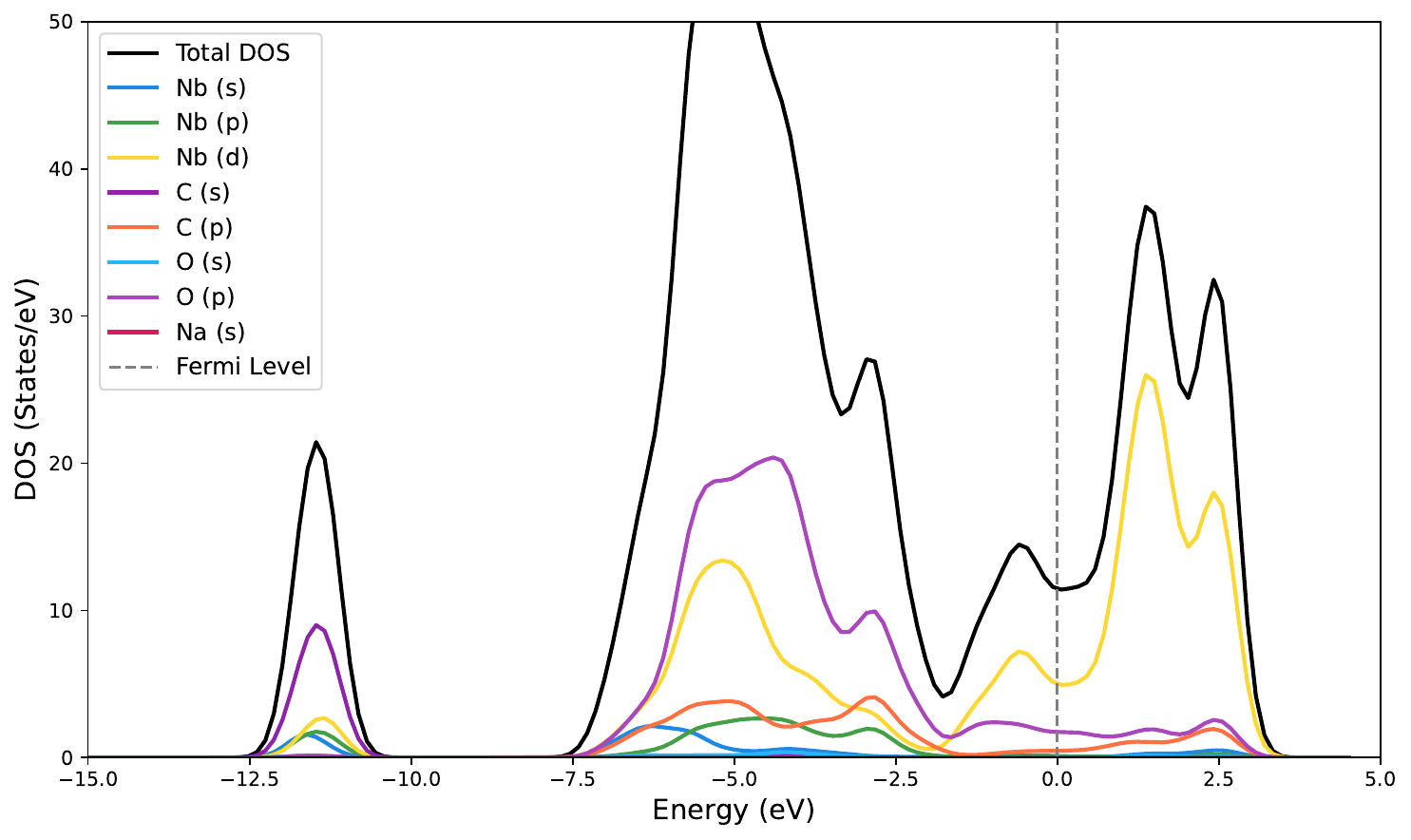}
  \end{subfigure}
  \begin{subfigure}{0.48\textwidth}
    \centering
    \caption*{\raisebox{1.5\height}{\hspace{40mm}\textbf{d) \NbCO{2}Li}}} % Adjust position
    \vspace{-3.75mm}
    \includegraphics[width=\textwidth]{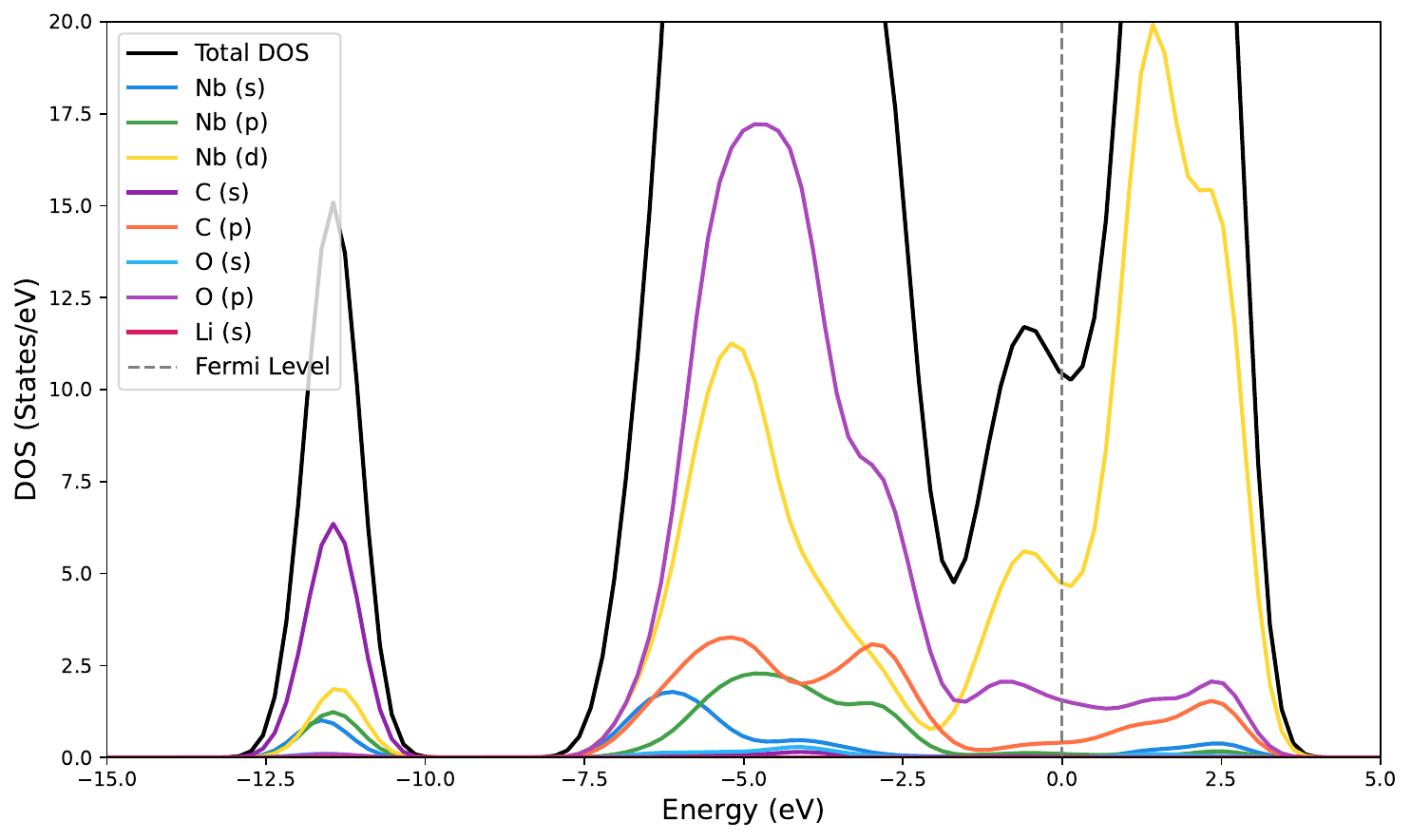}
  \end{subfigure}
  \caption{a) Total and projected density of states of a) \NbC{2}Na, b) \NbCO{2}Na, c) \NbC{2}Li and d) \NbCO{2}Li.}
  \label{fgr:projected_electronic_band_structure}
\end{figure*}

We computed the partial density of states (PDOS) to examine the electronic properties of \NbC{2} and \NbCO{2} in both pristine and metal-intercalated forms (Fig. \ref{fgr:electronic_band_structure} and \ref{fgr:projected_electronic_band_structure}).
% \todo[inline]{finish incomplete sentence in caption of \ref{fgr:electronic_band_structure}}
In \NbC{2}, with or without metal intercalation, Nb(d) and C(p) states dominate the upper valence band. In \NbCO{2}, oxygen functionalization shifts the dominant contribution to O(p), followed by Nb(d) and C(p), demonstrating the strong impact of the functional group O on the density of states near the Fermi level. 

Across all systems, Nb(d) states contribute the most to the conduction band. The electronic band structure analysis confirms the material's metallic nature, as multiple energy bands intersect the Fermi level (E = 0 eV), ensuring partially filled states for electron transport. PDOS analysis further shows that Nb(d) orbitals primarily drive the metallic properties, with a prominent peak near the Fermi level indicating a high carrier density. This high carrier concentration enhances charge conduction, reinforcing the strong metallic character of all the studied systems in the present study. Previous studies have linked such features to desirable performance characteristics in ion batteries, increasing the conductivity due to enhanced charge transfer between Mxene and the metal atoms\cite{liTheoreticalInvestigationMo2C2024}.

\subsubsection{Open Circuit Voltage}

We additionally calculated the open circuit voltage to assess the efficacy of the anode material using the following equation:

\begin{equation}
  \NbC{2}(O\textsubscript{2}) + xM \rightarrow \NbC{2}(O\textsubscript{2})M_{x}
\end{equation}

\begin{equation}
  \Delta G_f = E_f + P \Delta V_f - T \Delta S_f
\end{equation}

\begin{equation}
  OCV = -\frac{\Delta G_f}{x} \approx -\frac{E_{\NbC{2}(O\textsubscript{2})_{M_x}}-E_{\NbC{2}(O\textsubscript{2})}-xE_M}{x}
\end{equation}

\noindent At room temperature, the $P\Delta V$ term (related to pressure-volume work) is typically very small for solid-state systems, and the $T\Delta S$ term (entropy contribution) does not significantly affect the voltage. These effects contribute less than 0.01 V (~25 mev) to the overall open-circuit voltage (OCV), making $\Delta G$ approximately equal to $\Delta E$ in practical calculations for battery and electrode materials\cite{fanComputationalDesignPromising2023}. We considered single-sided adsorption to compare the effects of Li and Na metals on the OCV in the \NbC{2} and \NbCO{2} systems.

\begin{figure}[h]
  \centering
    \includegraphics[width=\columnwidth]{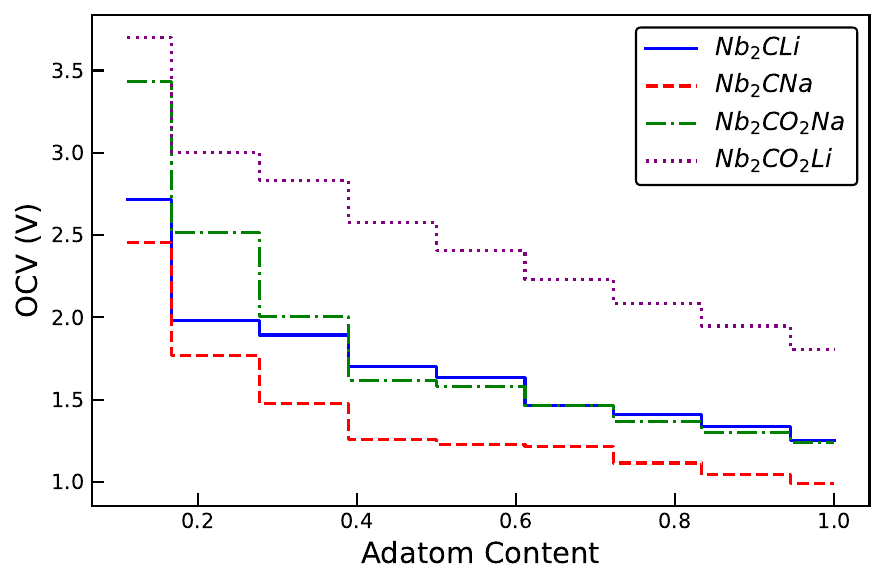}
    \caption{Calculated OCV changes for a) Na and b) Li on the substrate. The side views about the geometric structure of c, d) Na and e, f) Li adsorbed on \NbC{2}(O\textsubscript{2}).}
    \label{fgr:adatom_content}
\end{figure}

The OCVs of Li on \NbC{2}(O\textsubscript{2}) and Na on \NbC{2}(O\textsubscript{2}) show a decreasing trend with an increasing number of adatoms, as illustrated in Fig. \ref{fgr:adatom_content}. However, they exhibit different slopes, which could be attributed to variations in adsorption energetics, charge transfer mechanisms, and the interaction strength between the adatoms and the substrate. These factors influence how the OCV changes with increasing coverage, leading to distinct slope behaviors. The highest OCV was obtained for Li adsorption on \NbCO{2} at 3.70 eV, followed by 3.43 eV for Na adsorption on \NbCO{2}. In the \NbC{2} system, Li adsorption resulted in an OCV of 2.72 eV, while Na adsorption yielded a slightly lower value of 2.45 eV. Although Na consistently exhibited a lower OCV compared to Li, the difference remained relatively small, suggesting that Na can serve as a viable alternative. Additionally, the presence of oxygen functional groups on \NbC{2} significantly increased the OCV, emphasizing the role of surface chemistry in enhancing electrochemical performance. Given its comparable efficiency and lower cost, Na presents a promising alternative to Li for energy storage applications. We further calculated the theoretical capacity of Li/Na adsorbed \NbC{2} and \NbCO{2} systems. 

\begin{equation}
  C = \frac{nzF}{M}
\end{equation}

\noindent where $n$ is the concentration of adsorbed metal ions, $z$ is the valence state of metal ion, $F$ is Faraday constant (26801 mAh/mol). $M$ is the molar mass of \NbC{2}(O\textsubscript{2}).

\begin{table}[h]
  \small
  \centering
  \caption{\ Maximum Theoretical Gravimetric Capacity of Nb-Based MXenes for One-Sided and Two-Sided Li and Na Adsorption}
  \label{tbl:gravimetric_capacity}
  \begin{tabular*}{0.6\textwidth}{|@{\extracolsep{\fill}}l l c|}  % Use 'c' for centering
    \hline
    System & \makecell{Adsorption Site} & Theoritical Capacity ($mAh/g$)\\  
    \hline
    \multirow{2}{*}{\NbCLi{2}}  & Single Side & 130.89 \\
                                & Both Side & 261.78 \\
    \hline
    \multirow{2}{*}{\NbCNa{2}}  & Single Side & 121.37 \\
                                & Both Side & 242.74 \\
    \hline
    \multirow{2}{*}{\NbCO{2}Li} & Single Side & 113.20 \\
                                & Both Side & 226.40 \\
    \hline
    \multirow{2}{*}{\NbCO{2}Na} & Single Side & 106.01 \\
                                & - & - \\
    \hline
  \end{tabular*}
\end{table}

The theoretical gravimetric capacities of Nb-based MXenes for Li and Na adsorption were calculated considering both single-sided and double-sided adsorption. The results indicate a general trend where Li-based systems exhibit higher capacities than their Na-based counterparts due to the lower atomic weight of Li. However, the difference in capacity between Li and Na adsorption is relatively small, suggesting that Na could be a viable alternative to Li for energy storage applications.

For \NbC{2}-based systems, the highest theoretical capacity was obtained for \NbCLi{2}, with values of 130.89 mAh/g for single-sided adsorption and 261.78 mAh/g for double-sided adsorption. These results are closely aligned with previous studies on Li ion batteries, confirming the consistency of our calculations with established theoretical predictions\cite{santoy-floresNb2Nb2CO22024}. In contrast, \NbCNa{2} showed slightly lower capacities of 121.37 mAh/g (single-sided) and 242.74 mAh/g (double-sided), indicating that Na adsorption results in a minor capacity reduction.

When oxygen functionalization is introduced in \NbCO{2}-based systems, a notable decrease in capacity is observed. \NbCO{2}Li exhibits 113.20 mAh/g (single-sided) and 226.40 mAh/g (double-sided), whereas \NbCO{2}Na has the lowest capacities among all tested systems, with 106.01 mAh/g for single-sided adsorption. This trend suggests that the presence of oxygen functional groups influences charge storage by altering the adsorption environment, potentially affecting charge transfer and ion diffusion. Previous studies on different MXenes have also reported a similar reduction trend when an oxygen functional group is added to the metal carbide\cite{liEffectSfunctionalizedVacancies2020}. Oxygen functionalization in \NbCO{2} reduces interlayer spacing, increasing the diffusion barrier and restricting ion mobility leading to higher energy requirements for ion migration, ultimately lowering charge storage capacity compared to \NbC{2}.

\section*{Conclusions}%\todo{I think 3 or 5 bullet points would be better. response: Rather than adding bullet points, I have summerized them separate paragraphs.}%
In this study, we conducted an ab initio investigation of \NbC{2} and \NbCO{2} MXenes as potential anode materials for sodium- and lithium-ion batteries, focusing on their structural, electronic, and electrochemical properties. Our Raman analysis indicates that Li/Na intercalation predominantly alters the electronic environment rather than inducing significant structural distortions, as evidenced by intensity variations without notable peak shifts. Adsorption energy calculations confirm that the T4 and H3 sites are the most favorable for metal intercalation onto \NbC{2} and \NbCO{2} respectively. \NbCO{2} exhibits stronger adsorption due to the presence of O$_2$ functional groups irrespective of Li/Na intercalation.

The diffusion barrier analysis reveals a trade-off between adsorption stability and ion mobility. \NbCO{2} provides stronger adsorption sites for Li and Na, which enhances charge retention but also increases migration resistance, particularly along the upper-layer pathway (Path 2). In contrast, \NbC{2} exhibits lower diffusion barriers, especially for Na ions, making it a more favorable candidate for fast ion transport. This suggests that Na-ion batteries could achieve higher efficiency and faster charge/discharge rates compared to their Li-ion counterparts.

Bader charge analysis highlights that oxygen functionalization significantly enhances charge transfer, stabilizing intercalated ions and reinforcing electrostatic interactions. However, this increased charge localization also leads to higher energy barriers for ion migration. Electronic structure calculations confirm the metallic nature of \NbC{2} and \NbCO{2}, with Nb(d) orbitals contributing dominantly to the conduction band, ensuring efficient electron transport necessary for battery applications. Open circuit voltage (OCV) calculations further demonstrate that \NbCO{2} enhances electrochemical performance due to its higher OCV values, emphasizing the role of surface chemistry in modulating energy storage characteristics.

Theoretical capacity calculations show that while Li-based systems generally exhibit higher capacities due to Li's lower atomic weight, Na-based systems provide comparable performance, reinforcing their potential as a viable alternative. The presence of oxygen functional groups in \NbCO{2} reduces capacity due to increased diffusion resistance, highlighting the need for structural modifications to balance adsorption strength with ion transport kinetics.

Overall, our findings suggest that \NbC{2} and \NbCO{2} MXenes hold strong potential as anode materials for sodium- and lithium-ion batteries, with \NbC{2} being more suitable for applications requiring faster ion transport and \NbCO{2} offering enhanced charge retention. Given the lower cost and abundance of Na, sodium-ion batteries based on these MXenes present a promising alternative to conventional lithium-ion batteries. Future work should explore surface engineering strategies to optimize diffusion pathways and enhance electrochemical performance, enabling the development of high-performance, cost-effective energy storage solutions.

\section*{Author contributions}
\textbf{Nishat Sultana:} Conceptualization, Methodology, Formal Analysis, Writing-Original Draft, Writing-Review \& Editing. \textbf{Abdullah Al Amin:} Resources, Conceptualization, Writing-Original Draft, Writing-Review \& Editing. \textbf{Eric J. Payton:} Supervision, Writing-Original Draft, Manuscript Revision\& Editing. \textbf{Woo Kyun Kim:} Resources, Manuscript Revision 

\section*{Conflicts of interest}
The authors declare no conflicts of interest.

\section*{Data availability}
Data is available upon reasonable request.

\section*{Acknowledgements}

The authors acknowledge gracious support from the Ohio Supercomputer Center for the high performance computation resources through academic project.

%Bibliography
\bibliographystyle{unsrt}  
\bibliography{rsc}

\end{document}